\documentclass[journal,10pt]{IEEEtran}
\usepackage{pgfplots}
\usepackage{tikz}
\usetikzlibrary{calc}
\makeatletter
\newcommand{\gettikzxy}[3]{%
  \tikz@scan@one@point\pgfutil@firstofone#1\relax
  \edef#2{\the\pgf@x}%
  \edef#3{\the\pgf@y}%
}
\usetikzlibrary{spy,backgrounds}

\usepackage{acronym}
\acrodef{6g}[6G]{the sixth generation}
\acrodef{bs}[BS]{base station}
\acrodef{bse}[BSE]{beam squint effect}
\acrodef{cp}[CP]{cyclic prefix}
\acrodef{dof}[DOF]{degrees-of-freedom}
\acrodef{elaa}[ELAA]{extremely large antenna array}
\acrodef{ff}[FF]{far-field}
\acrodef{las}[L\&S]{localization and sensing}
\acrodef{los}[LOS]{line-of-sight}
\acrodef{nf}[NF]{near-field}
\acrodef{ris}[RIS]{reconfigurable intelligent surface}
\acrodef{rtt}[RTT]{round-trip-time}
\acrodef{sns}[SNS]{spatial non-stationarity}
\acrodef{swm}[SWM]{spherical wave model}

\acrodef{siso}[SISO]{single-input-single-output}
\acrodef{mimo}[MIMO]{multi-input-multi-output}
\acrodef{ue}[UE]{user equipment}
\acrodef{dmimo}[D-MIMO]{distributed MIMO}
\acrodef{sp}[SP]{scatter point}
\acrodef{awgn}[AWGN]{additive white Gaussian noise}
\acrodef{nlos}[NLOS]{non-line-of-sight}
\acrodef{ofdm}[OFDM]{orthogonal frequency division multiplexing}
\acrodef{tdoa}[TDOA]{time-difference-of-arrival}
\acrodef{toa}[TOA]{time-of-arrival}
\acrodef{am}[AM]{artificial multipath}
\acrodef{an}[AN]{artificial noise}
\acrodef{csi}[CSI]{channel state information}
\acrodef{mcrb}[MCRB]{misspecified Cramér-Rao bound}
\acrodef{crb}[CRB]{Cramér-Rao bound}
\acrodef{lb}[LB]{lower bound}
\acrodef{rmse}[RMSE]{root mean squared error}
\acrodef{psd}[PSD]{power spectral density}
\acrodef{pdf}[PDF]{probability distribution function}
\acrodef{aoa}[AOA]{angle-of-arrival}
\acrodef{aod}[AOD]{angle-of-departure}
\acrodef{fim}[FIM]{Fisher information matrix}
\acrodef{crb}[CRB]{Cramér-Rao bound}
\acrodef{moo}[MOO]{multi-objective optimization}
\acrodef{qos}[QoS]{quality of service}
\acrodef{sdp}[SDP]{semi-definite programming}
\acrodef{lmi}[LMI]{linear matrix inequality}
\acrodef{sdr}[SDR]{semi-definite relaxation}
\acrodef{rcs}[RCS]{radar cross section}
\acrodef{isac}[ISAC]{integrated sensing and communication}
\acrodef{bp}[BP]{bistatic positioning}
\acrodef{ms}[MS]{monostatic sensing}
\acrodef{slam}[SLAM]{simultaneous localization and mapping}
\acrodef{ms}[MS]{monostatic sensing}
\acrodef{fdb}[FDB]{full-dimensional beamforming}
\acrodef{cpa}[CPA]{codebook-based power allocation}
\acrodef{apa}[APA]{average power allocation}
\acrodef{qcqp}[QCQP]{quadratically constrained quadratic program}
\acrodef{dac}[DAC]{digital-to-analog converter}
\acrodef{ps}[PS]{phase shifter}
\acrodef{ao}[AO]{alternating optimization}
\acrodef{sqp}[SQP]{sequential quadratic programming}

\usepackage{booktabs} % for better table formatting
\usepackage{amsmath}
\usepackage{graphicx} %use graph format
\usepackage{epstopdf}
\usepackage{amssymb}
\usepackage{amsfonts}
\usepackage{amsthm}
\usepackage{cite}
\usepackage{bm,comment}
\usepackage{algorithm}
\usepackage{algpseudocode}

\usepackage{subeqnarray}
\usepackage{subfigure}
\usepackage{multicol}
\usepackage{multirow}
\usepackage{diagbox}
\usepackage{slashbox}
\usepackage{stfloats}
\usepackage{float}
\usepackage{color} 
\usepackage{cases}
\usepackage{lipsum}

\newtheorem{remark}{\bf{Remark}}

\usepackage{geometry}
\usepackage[colorlinks,linkcolor=blue,citecolor=red]{hyperref}
\allowdisplaybreaks
\begin{document}
\setlength{\textfloatsep}{4pt}

\bstctlcite{IEEEexample:BSTcontrol}
\title{Joint Bistatic Positioning and Monostatic Sensing: Optimized Beamforming and Performance Tradeoff}

\author{Yuchen Zhang, Hui Chen, Pinjun Zheng, Boyu Ning, Hong Niu, \\
Henk Wymeersch, \emph{Fellow, IEEE},
and Tareq Y. Al-Naffouri, \emph{Fellow, IEEE}

% \\
% \IEEEauthorrefmark{1}King Abdullah University of Science and Technology, KSA
% \ \ \ 
% \IEEEauthorrefmark{2}Chalmers University of Technology, Sweden
% \\
% \IEEEauthorrefmark{3}University of Electronic Science and Technology of China, China
% \\
% (yuchen.zhang@kaust.edu.sa)
% \thanks{

% }
\thanks{
This work was supported in part by the King Abdullah University of Science and Technology (KAUST) Office of Sponsored Research (OSR) under Award ORA-CRG2021-4695, and by the SNS JU project 6G-DISAC under the EU’s Horizon Europe research, by innovation programme under Grant Agreement No 101139130, and by the Swedish Research Council through the project HAILS under VR Grant 2022-03007.

Yuchen Zhang, Pinjun Zheng, and Tareq Y. Al-Naffouri are with the Electrical and Computer Engineering Program, Computer, Electrical and Mathematical Sciences and Engineering (CEMSE), King Abdullah University of Science and Technology (KAUST), Thuwal 23955-6900, Kingdom of Saudi Arabia (e-mail: \{yuchen.zhang; pinjun.zheng; tareq.alnaffouri\}@kaust.edu.sa).

Hui Chen and Henk Wymeersch are with the Department of Electrical Engineering, Chalmers University of Technology, 41296 Gothenburg, Sweden (e-mail: \{hui.chen; henkw\}@chalmers.se).

Boyu Ning is with the National Key Laboratory of Wireless Communications, University of Electronic Science and Technology of China, Chengdu 611731, China (e-mail: boydning@outlook.com). 

Hong Niu is with the School of Electrical and Electronics Engineering, Nanyang Technological University, Singapore 639798 (e-mail: hong.niu@ntu.edu.sg).
}
}
\maketitle
\begin{abstract}
We investigate joint bistatic positioning (BP) and monostatic sensing (MS) within a multi-input multi-output orthogonal frequency-division system. Based on the derived Cramér-Rao Bounds (CRBs), we propose novel beamforming optimization strategies that enable flexible performance trade-offs between BP and MS. Two distinct objectives are considered in this multi-objective optimization problem, namely, enabling user equipment to estimate its own position while accounting for unknown clock bias and orientation, and allowing the base station to locate passive targets. 
% This results in a multi-objective optimization problem for joint BP and MS via beamforming optimization. 
We first analyze digital schemes, proposing both weighted-sum CRB and weighted-sum mismatch (of beamformers and covariance matrices) minimization approaches. These are examined under full-dimension beamforming (FDB) and low-complexity codebook-based power allocation (CPA). To adapt to low-cost hardwares, we develop unit-amplitude analog FDB and CPA schemes based on the weighted-sum mismatch of the covariance matrices paradigm, solved using distinct methods. 
Numerical results confirm the effectiveness of our designs, highlighting the superiority of minimizing the weighted-sum mismatch of covariance matrices, and the advantages of mutual information fusion between BP and MS.
% Numerical results confirm the effectiveness of our designs, highlighting: 1) the superiority of minimizing the weighted-sum mismatch of covariance matrices, and 2) the advantages of mutual information fusion between BP and MS, offering valuable insights for practical system design.
\end{abstract}

\IEEEpeerreviewmaketitle

\section{Introduction}
\Ac{isac} represents one of the most transformative shifts in 6G networks, merging sensing and communication capabilities and exploiting the mutualistic mechanism to enable a wide range of novel applications\cite{fan2020tcom,fan2022jsac,andrew2021st,an2022st,shihang2024jiot,nuria2024proc}. Sensing, in this context, refers to a network's ability to detect, locate, and interpret information about objects or users within its environment, facilitating applications that leverage both communication and sensing information, such as autonomous navigation, environmental monitoring, and location-based/aware services\cite{nuria2024proc,hui2022st,behravan2022positioning,henk2024st-i,henk2024st-ii}. 
% While sensing has been long regarded as a byproduct in networks primarily designed for communications\cite{hui2022st,behravan2022positioning,nuria2024proc}, we are now witnessing a paradigm shift in which sensing has been anticipated being a native functionality and key enabler driving the development towards 6G and beyond. 
In general, sensing can be classified into sensing connected devices (e.g., via time-of-arrival-based positioning) and passive objects (e.g., via mono-/bi-/multi-static sensing)~\cite{behravan2022positioning,shihang2024jiot,nuria2024proc}.
% In general, sensing can be implemented in both monostatic and bistatic modes. In the monostatic mode, the transmitter and receiver are co-located, whereas in the bistatic mode, they are spatially separated \cite{an2022st,shihang2024jiot,nuria2024proc}. 
These two paradigms differ in their hardware and algorithmic requirements but can complement each other to enhance the network’s overall sensing capability\cite{yu2023globecom,nuria2024proc}.

Positioning of a connected device is widely adopted in current systems with the support of Global Navigation Satellite Systems and, more recently, cellular networks to support in urban and suburban areas where satellite visibility is often limited\cite{henk2022cl,hui2022st}. Specifically, the specification of positioning in 4G was introduced with the 3rd Generation Partnership Project (3GPP) Release 9, aiming to meet the regulatory requirement of 50-meter accuracy a \ac{ue}. The potential of positioning in 5G has been evaluated since 3GPP Release 15, and more ambitious efforts to achieve ultra-high accuracy, in conjunction with \ac{isac} for 6G, have been explored in Releases 18, 19, and beyond \cite{cha20245g,henk2024st-ii,nuria2024proc}.
The evolution of communication systems has attracted considerable research attention in recent years. However, positioning connected devices requires the target to be part of the network that can transmit and receive pilot signals, and the sensing of passive objects is largely ignored.

\Ac{ms} originates from radar technology, which has been widely used for military and civilian air surveillance\cite {olsen2017bridging}. In modern networks, especially with the higher frequencies anticipated in 6G, \ac{ms} allows \acp{bs} to act as multi-functional nodes, combining communication with radar-like sensing capabilities\cite{nuria2024proc,fan2023spm}. The expanded array apertures and bandwidths available in high-frequency bands, such as millimeter wave and sub-terahertz, significantly enhance the spatial and temporal resolution of estimation and hence the environment sensing performance~\cite{hui2022st}.
% establishing these bands as practical tools for detailed environmental sensing\cite{hui2022st}. 
Leveraging these capabilities, \ac{isac} has demonstrated substantial potential in enabling perceptive mobile networks\cite{fuwang2023twc,yonina2023wcm} and facilitating predictive beamforming in high-mobility scenarios\cite{liu2020radar,weijie2021twc}. Although \ac{ms} in communication systems has yet to be standardized, the International Telecommunication Union's Radio Communication Division technical report identifies \ac{isac} as a primary usage scenario, underscoring the indispensable role of \ac{ms} in its implementation \cite{kaushik2024toward}. 

In contrast, bistatic and multistatic sensing does not require full-duplex capability at the anchor, allowing for spatial diversity and extended coverage, which benefits various network applications\cite{chen2023riss}. All these mentioned techniques can sense passive targets, complementing positioning functions.
Considering the dynamic characteristics of the network, especially when the mobile users are part of the sensing tasks of passive objects, simultaneously positioning active devices and mapping environmental targets within a bistatic setup~\cite{henk2018twc,henk2019twc,ge2022computationally,yu2023globecom,nazari2023mmwave}, which is termed as \ac{bp} in this work. 
However, challenges such as the need for precise synchronization and orientation management between the transmitter and receiver must be overcome\cite{henk2024st-ii}.

% Despite the improved communication capacity and enhanced sensing accuracy, the introduced path loss necessitates beamforming gain to be achieved through large (electronically) controlled antenna arrays.
% As 6G technology evolves, the use of higher frequency bands offers improved communication capacity and enhanced sensing accuracy but also introduces significant path loss, highlighting the necessity for high beamforming gain achievable through large (electronically) controlled antenna arrays. 
Besides sensing modes and scenarios, beamforming optimization has been a key area of research in \ac{isac}, enhancing both sensing and communication performance and enabling effective tradeoffs between them. 
To balance these objectives, a common design criterion is to approximate an ideal \ac{mimo} radar beampattern while meeting communication performance requirements through beamforming optimization \cite{fan2018twc, fan2018tsp, xiang2020tsp, xiang2022jsac}. Extending this approach, recent studies \cite{fan2022tsp, xianxin2023tsp} have employed the \ac{crb} to quantify sensing performance, allowing a more precise characterization of the sensing-communication tradeoff in \ac{isac} systems. Furthermore, beamforming design in \ac{isac} has progressed from transmitter-only configurations to integrated transceiver designs \cite{chen2022jsac, rang2022jsac, zack2024tcom}.

Most of the aforementioned works focus on optimized beamforming for the efficient integration of radar-like \ac{ms} and communications, while overlooking beamforming design for \ac{bp}, which plays an increasingly important role in the generational upgrades of cellular networks \cite{nuria2024proc}. The authors of \cite{furkan2022tvt} examined the \ac{bp} setup with clock bias, shedding light on the properties of optimal beamforming. In \cite{henk2022jstsp}, the beamforming design for \ac{bp} was extended to the reconfigurable intelligent surface (RIS)-aided scenarios, enabling efficient joint \ac{bs}-RIS beamforming for improved \ac{bp} performance. As \ac{isac} advances toward 6G, both \ac{bp} and \ac{ms} are expected to coexist, and it is crucial to understand the tradeoff between these two paradigms to achieve complementary strengths. 
% This makes it crucial to understand the tradeoff between these two paradigms. 
However, it should be noted that \ac{bp} and \ac{ms} are typically studied independently, with limited attention given to their coexistence. The authors of \cite{yu2023globecom} initiated research on integrating \ac{bp} and \ac{ms} from a simultaneous localization and
mapping perspective. \emph{However, no existing works have designed beamformers to balance the tradeoff between these two paradigms.}

In this paper, we explore the joint tasks of \ac{bp} and \ac{ms} within a representative \ac{mimo} \ac{ofdm} framework, proposing effective beamforming designs that enable a flexible performance tradeoff between \ac{bp} and \ac{ms}, as characterized by the \ac{crb}. Our key contributions are summarized as follows.
\begin{itemize}
    % \item We systematically derive the CRBs for \ac{bp} and \ac{ms} as functions of beamformers, addressing two distinct objectives: enabling the \ac{ue} to estimate its own position (with unknown clock bias and orientation) and allowing the \ac{bs} to sense and position passive targets, including the UE.
    \item To optimize \ac{bp} and \ac{ms} jointly and strike a tradeoff between \ac{bp} and \ac{ms}, we formulate a \ac{moo} problem for beamforming design. 
    \begin{itemize}
        \item Starting with \emph{digital}\footnote{The term \emph{digital} beamforming in this paper refers to the ability to transmit amplitude-scaled and phase-shifted versions of a signal across multiple antennas, which are also referred to as analog active phased arrays with controllable per-antenna amplitude is sufficient \cite{ActiveAnalogArray}. In contrast, \emph{analog} beamforming typically relies on standard analog passive arrays without per-antenna amplitude control, which will also be examined in this work.
        % In contrast, \emph{analog} beamforming typically relies on standard analog passive arrays without per-antenna amplitude control, which will also be examined in this work. For simplicity, we refer to the former as \emph{digital} throughout the paper.
        } beamforming schemes, a weighted-sum \ac{crb} approach is proposed and solved using the \ac{fdb} method to ensure a weak Pareto frontier. This reveals the optimal beamforming structure, which in turn leads to a low-complexity \ac{cpa} method. Additionally, weighted-sum mismatch minimization approaches, commonly used in balance-pursuit problems, are introduced under two distinct paradigms: beamformer mismatch and covariance matrix mismatch. These approaches are solved using both the \ac{fdb} and \ac{cpa} methods.
        \item As hardware-efficient alternatives, \emph{analog} beamforming schemes are proposed based on the weighted-sum mismatch of covariance matrices. Using an \ac{ao} framework, we propose an \ac{fdb} method, where each alternation is solved through \ac{sqp}. Subsequently, we introduce an analog \ac{cpa} method based on analog codebook construction.
    \end{itemize}
    \item Comprehensive numerical results are presented to validate the effectiveness of the proposed beamforming schemes and to reveal the fundamental tradeoff between \ac{bp} and \ac{ms}. Specifically, we highlight the advantage of minimizing the weighted-sum mismatch of covariance matrices for beamforming, as it approaches the performance frontier achieved by the weighted-sum \ac{crb} approach. This finding supports the adoption of this paradigm when designing analog schemes. Furthermore, the results showcase the significant benefits of fusing common information between \ac{bp} and \ac{ms}, underscoring the importance of leveraging the mutualistic mechanism between \ac{bp} and \ac{ms} in practical system design.
\end{itemize}

The remainder of the paper is organized as follows. Section II introduces the system model and formulates the problem. Sections III and IV present digital beamforming designs based on the weighted-sum \ac{crb} and weighted-sum mismatch approaches, respectively. In Section V, we propose analog beamforming methods based on the weighted-sum mismatch of covariance matrices. Section VI provides an analysis of convergence and complexity, followed by numerical results in Section VII. Finally, Section VIII concludes the paper.

The primary notations used throughout this paper are defined as follows. Regular lowercase letters denote scalars, bold lowercase letters denote vectors, and bold uppercase letters represent matrices. The 2-norm of a vector $\mathbf{a}$ is denoted by $\left\|\mathbf{a}\right\|$, while the Frobenius norm of a matrix $\mathbf{A}$ is denoted by $\left\|\mathbf{A}\right\|_{\text{F}}$. Superscripts $\mathsf{T}$ and $\mathsf{H}$ indicate the transpose and Hermitian transpose of a vector or matrix, respectively. Additionally, $\text{tr}(\mathbf{A})$ and $\text{rank}(\mathbf{A})$ denote the trace and rank of matrix $\mathbf{A}$, and $\mathbf{A} \succeq \mathbf{0}$ indicates that matrix $\mathbf{A}$ is Hermitian and positive semi-definite. The real and imaginary parts of a scalar $a$ are represented by $\Re \left\{a\right\}$ and $\Im \left\{a\right\}$, respectively. $\text{diag}(\mathbf{a})$ represents a diagonal matrix with elements of $\mathbf{a}$ on its diagonal. Lastly, $\mathcal{CN}(\boldsymbol{\mu}, \mathbf{C})$ denotes a circularly symmetric complex Gaussian (CSCG) distribution with mean $\boldsymbol{\mu}$ and covariance matrix $\mathbf{C}$.

% \textbf{Beam design and tradeoffs}:
% Integrating both monostatic and bistatic sensing into a single system provides a more comprehensive sensing framework, enhancing both positioning accuracy and environmental awareness. However, combining these two modes introduces challenges, particularly in balancing the beamforming requirements for each mode. Monostatic and bistatic sensing differ in their signal and coverage requirements, which makes the joint beam design and resource allocation a complex task. Effective beamforming and carefully optimized tradeoff between sensing and communication performance are therefore essential for the success of ISAC in future networks.
% \red{More content and references are needed for these two paragraphs.}

% \textbf{Optimization methods}:
% Optimizing beamforming for ISAC requires multi-objective approaches that can simultaneously address the needs of both communication and sensing. Techniques such as Cramér-Rao Bound (CRB) minimization and multi-objective optimization (MOO) frameworks have been explored to tackle the distinct requirements of monostatic and bistatic sensing. These optimization methods ensure that beamforming designs can flexibly balance the tradeoff between achieving high communication rates and precise sensing, making them central to the development of ISAC-enabled 6G systems.

\begin{figure}[t]
\centering
\includegraphics[width=0.75\linewidth]{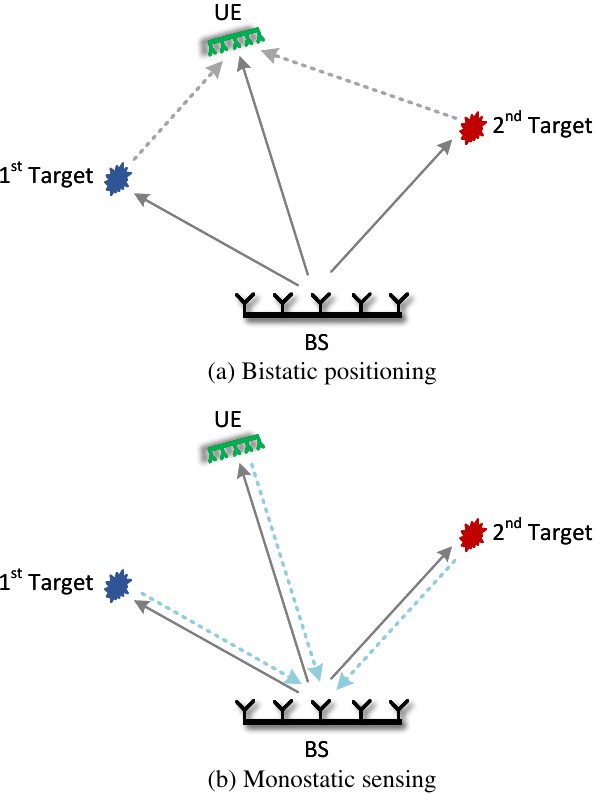}
\caption{Illustration of joint (a) BP and (b) MS, where the BS transmits pilot signals, functioning as a monostatic radar to sense passive targets and the UE. Meanwhile, the UE uses the received pilot signals to position itself.}\label{sys_mod}
\end{figure}

\section{System Model And Problem Formulation} 
\subsection{Signal Model}
As illustrated in Fig. \ref{sys_mod}, we consider a \ac{mimo} \ac{ofdm}-based joint \ac{bp} and \ac{ms} system with $M$ subcarriers, where a \ac{bs} equipped with $N_{\text{B}}$ transmit antennas transmits positioning pilot signals across $L$ slots to a \ac{ue} equipped with $N_{\text{U}}$ antennas, who uses the received signals to positioning itself, referred to as \ac{bp}. Meanwhile, the \ac{bs} acts as a monostatic radar with $N_{\text{B}}$ colocated receive antennas\footnote{Note that the number of receive antennas does not necessarily need to be equal to the number of transmit antennas. We set them equal merely to simplify the notation while the generalization is straightforward.}, sensing the environments by receiving echoes from passive targets and the \ac{ue}, then estimating their positions, referred to as \ac{ms}. In the system, a passive target in \ac{ms} creates one multipath in \ac{bp}.

Let $P$ denote the number of \ac{ofdm} pilot symbols in each slot. The transmit signal associated with the $p$-th symbol in the $l$-th slot over the $m$-th subcarrier is given by
\begin{equation}\label{tran-sig}
\boldsymbol{x}_{l,p,m} = \boldsymbol{f}_{l}s_{p,m},
\end{equation}
 where $\boldsymbol{f}_{l}\in \mathbb{C}^{N_{\text{B}}  }$ is the beamformer\footnote{To reduce the computational complexity of optimization, particularly in practical systems with a large number of subcarriers, we adopt the principle in \cite{furkan2022tvt,henk2022jstsp}, wherein a digital beamformer maintains coherence across subcarriers while being amplitude adjustable. This approach, though less flexible than standard digital beamforming techniques, strikes a balance between performance and computational efficiency.} for the $l$-th slot, and $s_{p,m}$ is the unit-modulus pilot symbol over the $m$-th subcarrier of the $p$-th symbol. 
 
 \subsubsection{Receive Signal at \ac{bp}}
 The signal received at the \ac{ue} is
\begin{equation}\label{re-sig-bi}
\overline{\boldsymbol{y}}_{l,p,m} = \boldsymbol{W}^{\mathsf{H}}\overline{\boldsymbol{H}}_m\boldsymbol{x}_{l,p,m} + \overline{\boldsymbol{z}}_{l,p,m},
\end{equation}
where $\boldsymbol{W}\in \mathbb{C}^{N_{\text{U}}\times N_{\text{U},\text{RF}}}$ is the analog combining matrix\footnote{To collect energy from all directions, we hereafter set $\boldsymbol{W} = \mathbf{I}_{N_{\text{U}}}$, which can equivalently be realized by an analog array at the \ac{ue} using a DFT codebook over $N_{\text{U}}\times L$ frames\cite{furkan2022tvt}.} at the \ac{ue}, with $N_{\text{U},\text{RF}}$ being the number of RF chains, $\overline{\boldsymbol{H}}_m \in \mathbb{C}^{N_{\text{U}}\times N_{\text{B}}}$ is the channel between the \ac{bs} and the \ac{ue} over the $m$-th subcarrier, given by
\begin{equation}\label{bi-chan}
\overline{\boldsymbol{H}}_m = \sum_{k = 0}^{K} \overline{\beta}_{k}e^{-\jmath 2 \pi m \Delta f \overline{\tau}_k} \boldsymbol{a}_{\text{U}}\left(\psi_k\right)  \boldsymbol{a}_{\text{B}}^{\mathsf{H}}\left(\theta_k\right),
\end{equation}
and $\overline{\boldsymbol{z}}_{l,p,m}\sim  \mathcal{CN}(\boldsymbol{0},\sigma^2 \boldsymbol{I}_{N_{\text{U}}})$ is the \ac{awgn} at the \ac{ue} receiver. Here, $\sigma^2 = F N_0 \Delta f$ is the noise power where $F$, $N_0$, and $\Delta f$ denote the noise figure, single-side \ac{psd}, and subcarrier spacing, respectively. 

In \eqref{bi-chan}, $K$ denotes the number of targets, and $\overline{\beta}_k$, $\overline{\tau}_k$, $\psi_k$, and $\theta_k$ are the complex channel gain, delay, \ac{aoa}, and \ac{aod}, respectively, associated with the $k$-th path. For notational convenience, the \ac{los} path of the channel is indexed by $k=0$. Specifically, $\psi_0$ and $\theta_0$ denote the \ac{aoa} and \ac{aod} with respect to the \ac{bs} and the \ac{ue}, respectively. Finally, $\boldsymbol{a}_{\text{B}}\left(\cdot\right) \in \mathbb{C}^{N_{\text{B}} }$ and $\boldsymbol{a}_{\text{U}}\left(\cdot\right) \in \mathbb{C}^{N_{\text{U}} }$ are the steering vectors at the \ac{bs} (transmitter side) and the \ac{ue}, respectively.

\subsubsection{Receive Signal at \ac{ms}}
Similarly, the signal received at the (colocated) \ac{bs} receiver is 
\begin{equation}\label{re-sig-mono}
\underline{\boldsymbol{y}}_{l,p,m} = \underline{\boldsymbol{H}}_m\boldsymbol{x}_{l,p,m} + \underline{\boldsymbol{z}}_{l,p,m},
\end{equation}
where $\underline{\boldsymbol{H}}_m \in \mathbb{C}^{N_{\text{B}}\times N_{\text{B}}}$ is the round-trip channel between the \ac{bs} and the passive targets (including the \ac{ue}) over the $m$-th subcarrier, given by
\begin{equation}\label{mono-chan}
\underline{\boldsymbol{H}}_m = \sum_{k = 0}^{K} \underline{\beta}_{k}e^{-\jmath 2 \pi m \Delta f \underline{\tau}_k} \boldsymbol{a}_{\text{B}}\left(\theta_k\right)  \boldsymbol{a}_{\text{B}}^{\mathsf{H}}\left(\theta_k\right),    
\end{equation}
and $\underline{\boldsymbol{z}}_{l,p,m}\sim  \mathcal{CN}(\boldsymbol{0},\sigma^2 \boldsymbol{I}_{N_{\text{B}}})$ is the \ac{awgn} at the \ac{bs} receiver. Here, $\underline{\beta}_k$ and $\underline{\tau}_k$ represent the complex channel gain and delay, respectively, associated with the $k$-th object. Here, \ac{ue} is also an target, indexed by $k=0$, in the \ac{ms} scenario.

\subsection{CRB-Based Performance Metric}
For both \ac{bp} and \ac{ms}, we consider a two-stage positioning process, where the channel-domain parameters are estimated in the first stage, followed by the inference of position-domain parameters in the second stage.

\subsubsection{Performance Metric of \ac{bp}}
In \ac{bp}, the channel-domain parameters are collected by $\overline{\boldsymbol{\xi}} = [\boldsymbol{\theta}^{\mathsf{T}},\boldsymbol{\psi}^{\mathsf{T}},\overline{\boldsymbol{\tau}}^{\mathsf{T}},\boldsymbol{\overline{\beta}}_{\text{R}}^{\mathsf{T}},\boldsymbol{\overline{\beta}}_{\text{I}}^{\mathsf{T}}]^{\mathsf{T}} \in \mathbb{R}^{(5K+5) }$, where $\boldsymbol{\theta} = [{\theta}_0, \ldots, {\theta}_K]^{\mathsf{T}} \in \mathbb{R}^{(K+1) }$ is the collection of \acp{aod}, $\boldsymbol{\psi} = [{\psi}_{0}, \ldots, {\psi}_{K}]^{\mathsf{T}} \in \mathbb{R}^{(K+1) }$ is the collection of \acp{aoa}, $\overline{\boldsymbol{\tau}} = [{\overline{\tau}}_{0}, \ldots, {\overline{\tau}}_{K}]^{\mathsf{T}} \in \mathbb{R}^{(K+1) }$ represents the delays, and $\boldsymbol{\overline{\beta}}_{\text{R}} = [\Re\{{\overline{\beta}}_{0}\}, \ldots, \Re\{{\overline{\beta}}_{K}\}]^{\mathsf{T}} \in \mathbb{R}^{(K+1) }$ and $\boldsymbol{\overline{\beta}}_{\text{I}} = [\Im\{{\overline{\beta}}_{0}\}, \ldots, \Im\{{\overline{\beta}}_{K}\}]^{\mathsf{T}} \in \mathbb{R}^{(K+1) }$ are the collections of the real and imaginary parts of the complex channel gains, respectively. Using the Slepian-Bangs formula\cite{furkan2022tvt}, the element at the $i$-th row and $j$-th column of the channel-domain \ac{fim} $\boldsymbol{I}_{\text{Chan}}(\overline{\boldsymbol{\xi}})$ is derived as
\begin{eqnarray}\label{bi-chan-fim}
\begin{aligned}
\left[\boldsymbol{I}_{\text{Chan}}\left(\overline{\boldsymbol{\xi}}\right)\right]_{i,j} &= \frac{2}{\sigma^2}\sum_{l=1}^{L}\sum_{p=1}^{P}
\sum_{m=1}^{M}\Re \left\{\frac{\partial \overline{\boldsymbol{\mu}}_{l,p,m}^{\mathsf{H}}}{\partial \left[\overline{\boldsymbol{\xi}}\right]_{i}}\frac{\partial \overline{\boldsymbol{\mu}}_{l,p,m}}{\partial \left[\overline{\boldsymbol{\xi}}\right]_{j}}\right\}\\
&= \frac{2N}{\sigma^2}\sum_{m=1}^{M}\Re \left\{\text{tr}\left(\frac{\partial \overline{\boldsymbol{H}}_{m}}{\partial \left[\overline{\boldsymbol{\xi}}\right]_{j}} \boldsymbol{F}\boldsymbol{F}^{\mathsf{H}}\frac{\partial \overline{\boldsymbol{H}}_{m}^{\mathsf{H}}}{\partial \left[\overline{\boldsymbol{\xi}}\right]_{i}}\right)\right\},
\end{aligned}
\end{eqnarray}
where $\overline{\boldsymbol{\mu}}_{l,p,m} = \overline{\boldsymbol{H}}_m\boldsymbol{x}_{l,p,m}$ denotes the noise-free observation from \eqref{re-sig-bi} and $\boldsymbol{F} = [\boldsymbol{f}_1,\ldots,\boldsymbol{f}_L] \in \mathbb{C}^{N_{\text{B}} \times L}$ collects $L$ beamformers.

The position-domain parameters are collected in $\overline{\boldsymbol{\eta}} = [\boldsymbol{p}_{\text{U}}^{\mathsf{T}},\Delta \phi,\boldsymbol{p}_{1}^{\mathsf{T}},\ldots,\boldsymbol{p}_{K}^{\mathsf{T}},\Delta t, \boldsymbol{\overline{\beta}}_{\text{R}}^{\mathsf{T}},\boldsymbol{\overline{\beta}}_{\text{I}}^{\mathsf{T}}]^{\mathsf{T}} \in \mathbb{R}^{(4K+6) }$, where $\boldsymbol{p}_{\text{U}} \in \mathbb{R}^{2  }$ represents the position of the \ac{ue}, and $\boldsymbol{p}_{k} \in \mathbb{R}^{2  }$ represents the position of the $k$-th object. The variable $\Delta \phi$ denotes the relative orientation of the \ac{bs} (in the \ac{ue}'s local coordinate system), while $\Delta t$ characterizes the clock bias that reflects the asynchronism between the \ac{bs} and \ac{ue} in the bistatic setting. Note that the nuisance parameters $\boldsymbol{\overline{\beta}}_{\text{R}}$ and $\boldsymbol{\overline{\beta}}_{\text{I}}$ from the channel-domain parameter $\overline{\boldsymbol{\xi}}$ remain part of the position-domain parameter $\overline{\boldsymbol{\eta}}$, as they do not contribute useful information for position estimation. Using the channel-domain \ac{fim}, the position-domain \ac{fim} $\boldsymbol{I}_{\text{Pos}}(\overline{\boldsymbol{\eta}})$ is computed as follows
\begin{equation}\label{bi-pos-fim}
\boldsymbol{I}_{\text{Pos}}\left(\overline{\boldsymbol{\eta}}\right) = \overline{\boldsymbol{J}}^{\mathsf{T}} \boldsymbol{I}_{\text{Chan}}\left(\overline{\boldsymbol{\xi}}\right)   \overline{\boldsymbol{J}},
\end{equation}
where $\overline{\boldsymbol{J}} \in  \mathbb{R}^{(5K+5)\times (4K+6)}$ is the Jacobian matrix, with the element in the $i$-th row and $j$-th column given by $[\overline{\boldsymbol{J}}]_{i,j} = \partial [\overline{\boldsymbol{\xi}}]_{i} / \partial [\overline{\boldsymbol{\eta}}]_{j}$. The \ac{crb} is used to quantify the \ac{bp} accuracy concerning $\boldsymbol{p}_{\text{U}}$, providing a lower bound on the sum of the covariances for estimating $\boldsymbol{p}_{\text{U}}$, and is expressed as
\begin{equation}\label{crb-pos-bi}
\overline{\text{CRB}}\left(\boldsymbol{p}_{\text{U}}\right) = \text{tr}\left(\left[\boldsymbol{I}_{\text{Pos}}\left(\overline{\boldsymbol{\eta}}\right)^{-1}\right]_{1 : 2, 1 : 2}\right).    
\end{equation}

\subsubsection{Performance Metric of \ac{ms}}
Following similar steps, the position-domain \ac{fim} for \ac{ms} is given by
\begin{equation}\label{mono-pos-fim}
\boldsymbol{I}_{\text{Pos}}\left(\underline{\boldsymbol{\eta}}\right) = \underline{\boldsymbol{J}}^{\mathsf{T}} \boldsymbol{I}_{\text{Chan}}\left(\underline{\boldsymbol{\xi}}\right)   \underline{\boldsymbol{J}},
\end{equation}
where $\underline{\boldsymbol{\xi}} = [\boldsymbol{\theta}^{\mathsf{T}},\boldsymbol{\underline{\tau}}^{\mathsf{T}},\boldsymbol{\underline{\beta}}_{\text{R}}^{\mathsf{T}},\boldsymbol{\underline{\beta}}_{\text{I}}^{\mathsf{T}}]^{\mathsf{T}} \in \mathbb{R}^{(4K+4) }$ and $\underline{\boldsymbol{\eta}} = [\boldsymbol{p}_{\text{U}}^{\mathsf{T}},\boldsymbol{p}_{1}^{\mathsf{T}},\ldots,\boldsymbol{p}_{K}^{\mathsf{T}},\boldsymbol{\underline{\beta}}_{\text{R}}^{\mathsf{T}},\boldsymbol{\underline{\beta}}_{\text{I}}^{\mathsf{T}}]^{\mathsf{T}} \in \mathbb{R}^{(4K+4) }$ are the channel-domain and position-domain parameters, respectively. Here, $\boldsymbol{\underline{\tau}} = [{\underline{\tau}}_{0}, \ldots, {\underline{\tau}}_{K}]^{\mathsf{T}} \in \mathbb{R}^{(K+1) }$ represents the delay measurements, while $\boldsymbol{\underline{\beta}}_{\text{R}} = [\Re\{{\underline{\beta}}_{0}\}, \ldots, \Re\{{\underline{\beta}}_{K}\}]^{\mathsf{T}} \in \mathbb{R}^{(K+1) }$ and $\boldsymbol{\underline{\beta}}_{\text{I}} = [\Im\{{\underline{\beta}}_{0}\}, \ldots, \Im\{{\underline{\beta}}_{K}\}]^{\mathsf{T}} \in \mathbb{R}^{(K+1) }$ represent the real and imaginary parts of the complex channel gains, respectively. 
% The element at the $i$-th row and $j$-th column of the channel-domain \ac{fim} $\boldsymbol{I}_{\text{Chan}}(\underline{\boldsymbol{\xi}})$ is derived as
% \begin{equation}\label{mono-chan-fim}
% \left[\boldsymbol{I}_{\text{Chan}}\left(\underline{\boldsymbol{\xi}}\right)\right]_{i,j} = \frac{2N}{\sigma^2}\sum_{m=1}^{M}\Re \left\{\text{tr}\left(\frac{\partial \underline{\boldsymbol{H}}_{m}}{\partial \left[\underline{\boldsymbol{\xi}}\right]_{j}} \boldsymbol{F}\boldsymbol{F}^{\mathsf{H}}\frac{\partial \underline{\boldsymbol{H}}_{m}^{\mathsf{H}}}{\partial \left[\underline{\boldsymbol{\xi}}\right]_{i}}\right)\right\},
% \end{equation}
% while the element in the $i$-th row and $j$-th column of the Jacobian matrix $\underline{\boldsymbol{J}} \in \mathbb{R}^{(4K+4)\times (4K+4)}$ given by $[\underline{\boldsymbol{J}}]_{i,j} = \partial [\underline{\boldsymbol{\xi}}]_{i} / \partial [\underline{\boldsymbol{\eta}}]_{j}$.
The \ac{crb} for \ac{ms}, concerning the passive targets (as well as the \ac{ue}), provides a lower bound on the sum covariance for estimating $\boldsymbol{p} = [\boldsymbol{p}_{\text{U}}^{\mathsf{T}},\boldsymbol{p}_{1}^{\mathsf{T}},\ldots,\boldsymbol{p}_{K}^{\mathsf{T}}]^{\mathsf{T}} \in \mathbb{R}^{(2K+2) }$ at the \ac{bs}, and is given by
\begin{equation}\label{crb-pos-mono}
\underline{\text{CRB}}\left(\boldsymbol{p}\right) = \text{tr}\left(\left[\boldsymbol{I}_{\text{Pos}}\left(\underline{\boldsymbol{\eta}}\right)^{-1}\right]_{1 : 2K + 2, 1 : 2K + 2}\right).    
\end{equation}

\subsection{Problem Formulation}
We observe that both $\overline{\text{CRB}}(\boldsymbol{p}_{\text{U}})$ and $\underline{\text{CRB}}(\boldsymbol{p})$ are functions of $\boldsymbol{F}$, which can be optimized by designing the beamformers $\boldsymbol{F}$\cite{furkan2022tvt,henk2022jstsp}. However, due to the different objectives, a performance tradeoff between \ac{bp} and \ac{ms} emerges. Specifically, this bistatic-monostatic performance tradeoff is characterized by a \ac{moo} problem\cite{ehrgott2005multicriteria}, expressed as
\begin{subequations}\label{moo-prob}
\begin{align}
\mathop {\min }\limits_{\boldsymbol{F}} \;\;\; &\left[\overline{\text{CRB}}\left(\boldsymbol{p}_{\text{U}}\right), \underline{\text{CRB}}\left(\boldsymbol{p}\right)\right]\label{moo-prob-obj}\\
{\rm{s.t.}}\;\;\;
&  \text{tr}\left(\boldsymbol{F}\boldsymbol{F}^{\mathsf{H}}\right) \le P_{\text{B}}/M, \label{moo-prob-power}
\end{align}
\end{subequations}
where $P_{\text{B}}$ is the power budget. Without loss of generality, the right-hand side of \eqref{moo-prob-power} is set as $P_{\text{B}}/M$ such that the total transmit power over $M$ subcarriers is $P$. Note that the optimal solution to \eqref{moo-prob} represents the Pareto frontier of $[\overline{\text{CRB}}(\boldsymbol{p}_{\text{U}}), \underline{\text{CRB}}(\boldsymbol{p})]$, which is challenging to find due to the \ac{moo} nature. Additionally, neither $\overline{\text{CRB}}(\boldsymbol{p}_{\text{U}})$ nor $\underline{\text{CRB}}(\boldsymbol{p})$ is convex with respect to $\boldsymbol{F}$, further complicating the problem. 

\begin{remark}\label{remark-fusion}
We would like to emphasize that in the previous formulation, the \ac{bp} and \ac{ms} components of the system are treated independently, with no exchange of information between them. However, it is important to recognize that, although positioning targets is not the primary goal of \ac{bp}, it remains a fundamental requirement, as does positioning the \ac{ue}. Therefore, despite operating in different configurations, both \ac{bp} and \ac{ms} share the common objective of jointly positioning the targets and the \ac{ue}. To understand the performance limits of joint \ac{bp} and \ac{ms}, we can leverage a mutualistic approach by combining the information from both components (assuming the existence of a feedback channel between the \ac{bs} and \ac{ue}), thus forming a bounding framework. Specifically, the fused position-domain parameters are aggregated as $\underline{\overline{\boldsymbol{\eta}}} = [\boldsymbol{p}_{1}^{\mathsf{T}},\ldots,\boldsymbol{p}_{K}^{\mathsf{T}},\boldsymbol{p}_{\text{U}}^{\mathsf{T}},\Delta \phi,\Delta t, \boldsymbol{\overline{\beta}}_{\text{R}}^{\mathsf{T}},\boldsymbol{\overline{\beta}}_{\text{I}}^{\mathsf{T}},\boldsymbol{\underline{\beta}}_{\text{R}}^{\mathsf{T}},\boldsymbol{\underline{\beta}}_{\text{I}}^{\mathsf{T}}]^{\mathsf{T}} \in \mathbb{R}^{(6K+6)}$. The elements of the fused position-domain \ac{fim}, denoted as $\boldsymbol{I}_{\text{Pos}}(\underline{\overline{\boldsymbol{\eta}}})$, are obtained by either replicating the exclusive terms from the position-domain \ac{fim} of \ac{bp} or \ac{ms}, or summing the relevant terms from both. The fused \ac{crb} for \ac{bp}, denoted as $\underline{\overline{\text{CRB}}}(\boldsymbol{p}_{\text{U}})$, and the fused \ac{crb} for \ac{ms}, denoted as $\underline{\overline{\text{CRB}}}(\boldsymbol{p})$, can be derived from $\boldsymbol{I}_{\text{Pos}}(\underline{\overline{\boldsymbol{\eta}}})$ by inverting it and extracting the appropriate trace terms. Additionally, although the beamforming methods we develop in the subsequent sections are based on the non-fused scenario, they can be extended to the fused case by solving a similar \ac{moo} problem as in \eqref{moo-prob} using the fused \acp{crb}.
\end{remark}

\section{Weighted-Sum CRB Optimization}
To address \eqref{moo-prob} and explore the performance tradeoff, we begin by applying the weighted-sum method, a well-established technique for obtaining the weak Pareto frontier of \ac{moo} problems\cite{ehrgott2005multicriteria}. Using this method, we formulate and solve an \ac{fdb} optimization problem aimed at optimizing the beamformers $\boldsymbol{F}$. Additionally, we reveal the characteristics of the optimal solution, which motivates the development of a \ac{cpa} approach that balances \ac{bp} and \ac{ms} with reduced complexity, serving as a complementary solution to the \ac{fdb} method.

\subsection{Full-Dimensional Beamforming}
With \ac{fdb}, we retain the beamformers $\boldsymbol{F}$ as the optimization variable. Under the weighted-sum approach, the problem in \eqref{moo-prob} is then reformulated as
\begin{subequations}\label{moo-prob-weight}
\begin{align}
\mathop {\min }\limits_{\boldsymbol{F}} \;\;\; &\alpha \overline{\text{CRB}}\left(\boldsymbol{p}_{\text{U}}\right) + \left(1 - \alpha\right)\underline{\text{CRB}}\left(\boldsymbol{p}\right)
\label{moo-prob-weight-obj}\\
{\rm{s.t.}}\;\;\;
&  \text{tr}\left(\boldsymbol{F}\boldsymbol{F}^{\mathsf{H}}\right) \le P_{\text{B}}/M, \label{moo-prob-weight-power}
\end{align}
\end{subequations}
where $\alpha \in [0, 1]$ is a constant that adjusts the priority between \ac{bp} and \ac{ms}, determined by the specific application scenario and \ac{qos} requirements.

Problem \eqref{moo-prob-weight} remains challenging to solve due to its non-convexity. By defining $\boldsymbol{V} = \boldsymbol{F}\boldsymbol{F}^{\mathsf{H}}$, we lift \eqref{moo-prob-weight} into a relaxed form (by omitting the constraint $\text{rank}(\boldsymbol{V}) = L$) as
\begin{subequations}\label{moo-prob-weight-sdr}
\begin{align}
\mathop {\min }\limits_{\boldsymbol{V}} \;\;\; &\alpha \overline{\text{CRB}}\left(\boldsymbol{p}_{\text{U}}\right) + \left(1 - \alpha\right)\underline{\text{CRB}}\left(\boldsymbol{p}\right)
\label{moo-prob-weight-sdr-obj}\\
{\rm{s.t.}}\;\;\;
&  \text{tr}\left(\boldsymbol{V}\right) \le P_{\text{B}}/M, \quad \boldsymbol{V} \succeq \mathbf{0}. \label{moo-prob-weight-sdr-power}
\end{align}
\end{subequations}
Next, note that the matrices on the right-hand sides of \eqref{crb-pos-bi} and \eqref{crb-pos-mono} can be reformulated as\cite{henk2019twc}
\begin{subequations}\label{efim}
\begin{align}
&\left[\boldsymbol{I}_{\text{Pos}}\left(\overline{\boldsymbol{\eta}}\right)^{-1}\right]_{1 : 2, 1 : 2} = \left[\overline{\boldsymbol{Y}} - \overline{\boldsymbol{G}}\overline{\boldsymbol{Z}}^{-1}\overline{\boldsymbol{G}}^{\mathsf{T}}\right]^{-1}, \\
&\left[\boldsymbol{I}_{\text{Pos}}\left(\underline{\boldsymbol{\eta}}\right)^{-1}\right]_{1 : 2K + 2, 1 : 2K + 2} = \left[\underline{\boldsymbol{Y}} - \underline{\boldsymbol{G}}\underline{\boldsymbol{Z}}^{-1}\underline{\boldsymbol{G}}^{\mathsf{T}}\right]^{-1},
\end{align}
\end{subequations}
where $\overline{\boldsymbol{Y}} = [\boldsymbol{I}_{\text{Pos}}(\overline{\boldsymbol{\eta}})]_{1 : 2, 1 : 2}$, $\overline{\boldsymbol{G}} = [\boldsymbol{I}_{\text{Pos}}(\overline{\boldsymbol{\eta}})]_{1 : 2, 3 : 4K + 6}$, $\overline{\boldsymbol{Z}} = [\boldsymbol{I}_{\text{Pos}}(\overline{\boldsymbol{\eta}})]_{3 : 4K + 6, 3 : 4K + 6}$, $\underline{\boldsymbol{Y}} = [\boldsymbol{I}_{\text{Pos}}(\underline{\boldsymbol{\eta}})]_{1 : 2K + 2, 1 : 2K + 2}$, $\underline{\boldsymbol{G}} = [\boldsymbol{I}_{\text{Pos}}(\underline{\boldsymbol{\eta}})]_{1 : 2K + 2, 2K + 3 : 4K + 4}$, and $\underline{\boldsymbol{Z}} = [\boldsymbol{I}_{\text{Pos}}(\underline{\boldsymbol{\eta}})]_{2K + 3 : 4K + 4, 2K + 3 : 4K + 4}$. 
By introducing auxiliary variables $\overline{\boldsymbol{U}}\in \mathbb{R}^{2 \times 2}$ and $\underline{\boldsymbol{U}}\in \mathbb{R}^{(2K+2)\times (2K+2)}$, \eqref{moo-prob-weight-sdr} can be reformulated into an equivalent form as
% \footnote{Due to space constraints, the detailed illustration of the equivalence between problems \eqref{moo-prob-weight-sdr} and \eqref{moo-prob-weight-relax} is omitted.}
\begin{subequations}\label{moo-prob-weight-relax}
\begin{align}
\mathop {\min }\limits_{\boldsymbol{V},\overline{\boldsymbol{U}},\underline{\boldsymbol{U}}} \;\;\; &\alpha \text{tr}\left(\overline{\boldsymbol{U}}^{-1}\right) + \left(1 - \alpha\right)\text{tr}\left(\underline{\boldsymbol{U}}^{-1}\right)
\label{moo-prob-weight-relax-obj}\\
{\rm{s.t.}}\;\;\;
& 
\begin{bmatrix}
\overline{\boldsymbol{Y}} - \overline{\boldsymbol{U}} & \overline{\boldsymbol{G}}\\
\overline{\boldsymbol{G}}^{\mathsf{T}} & \overline{\boldsymbol{Z}}
\end{bmatrix} \succeq \mathbf{0}, \quad
\begin{bmatrix}
\underline{\boldsymbol{Y}} - \underline{\boldsymbol{U}} & \underline{\boldsymbol{G}}\\
\underline{\boldsymbol{G}}^{\mathsf{T}} & \underline{\boldsymbol{Z}}
\end{bmatrix} \succeq \mathbf{0},\label{moo-prob-weight-relax-lmi}\\
& \overline{\boldsymbol{U}} \succeq \mathbf{0}, \quad \underline{\boldsymbol{U}} \succeq \mathbf{0}, \label{moo-prob-weight-relax-psd}\\
&  \text{tr}\left(\boldsymbol{V}\right) \le P_{\text{B}}/M,  \quad \boldsymbol{V} \succeq \mathbf{0}. 
\end{align}
\end{subequations}

The above problem is a convex \ac{sdp} problem that can be efficiently solved using off-the-shelf optimization tools such as CVX. Once solved, the beamformers $\boldsymbol{F}$ can be recovered from $\boldsymbol{V}$ via matrix decomposition or a randomization procedure\cite{tom2010spm}.

\begin{remark}\label{remark-opt-struct}
Following similar reasoning to that in Appendix C of \cite{jianli2008tsp}, we point out that the optimal covariance matrix that minimizes the \ac{crb} can be expressed as $\boldsymbol{V} = \boldsymbol{U}\boldsymbol{\Lambda}\boldsymbol{U}^{\mathsf{H}}$, where $\boldsymbol{\Lambda} \in \mathbb{C}^{\left(2K + 2\right) \times \left(2K + 2\right)}$ is a positive semi-definite matrix and $\boldsymbol{U} = [\boldsymbol{a}_{\text{B}}\left(\theta_0\right), \ldots, \boldsymbol{a}_{\text{B}}\left(\theta_K\right), \dot{\boldsymbol{a}}_{\text{B}}\left(\theta_0\right), \ldots, \dot{\boldsymbol{a}}_{\text{B}}\left(\theta_K\right)] \in \mathbb{C}^{N_{\text{B}} \times \left(2K + 2\right)}$, with $ \dot{\boldsymbol{a}}_{\text{B}}\left(\theta_0\right) = \partial \boldsymbol{a}_{\text{B}}\left(\theta_k\right)/\partial \theta_k$. 
It is important to note that although this property was derived under single \ac{crb} minimization (rather than weighted-sum minimization as in \eqref{moo-prob-weight-sdr}), the derivation in \cite{jianli2008tsp} can be straightforwardly extended to our case since the \ac{bp} and \ac{ms} components share the same $\boldsymbol{U}$. Therefore, it is omitted here for brevity. The revealed optimal structure of the solution can be applied to solve the \ac{sdp} problem in \eqref{moo-prob-weight-relax}, i.e., by optimizing $\boldsymbol{\Lambda}$ instead of $\boldsymbol{V}$, which significantly reduces the complexity. This is because the dimension of $\boldsymbol{\Lambda}$, determined by the number of targets, is typically much smaller than that of $\boldsymbol{V}$, which is determined by the number of transmit antennas.
\end{remark}

\subsection{Codebook-based Power Allocation}
By constraining $\boldsymbol{\Lambda}$ to be diagonal, the \ac{fdb} method is reduced to a lower-dimensional, lower-complexity power allocation scheme over the predetermined codebook matrix $\boldsymbol{U}$. Let $\boldsymbol{\rho} = [\rho_1,\ldots,\rho_{2K+2}]$ be the power allocation vector. We propose a \ac{cpa} method by substituting the variable $\boldsymbol{V}$ in \eqref{moo-prob-weight-sdr} with $\boldsymbol{U}\text{diag}\left(\boldsymbol{\rho}\right)\boldsymbol{U}^{\mathsf{H}}$. The resulting power allocation problem is then given by 
\begin{subequations}\label{moo-prob-weight-cpa}
\begin{align}
\mathop {\min }\limits_{\boldsymbol{\rho}} \;\;\; &\alpha \overline{\text{CRB}}\left(\boldsymbol{p}_{\text{U}}\right) + \left(1 - \alpha\right)\underline{\text{CRB}}\left(\boldsymbol{p}\right)
\label{moo-prob-weight-cpa-obj}\\
{\rm{s.t.}}\;\;\;
&  \text{tr}\left(\boldsymbol{U}\text{diag}\left(\boldsymbol{\rho}\right)\boldsymbol{U}^{\mathsf{H}}\right) \le \frac{P}{M}, \quad \boldsymbol{\rho} \geq \boldsymbol{0}, \label{moo-prob-weight-cpa-power}
\end{align}
\end{subequations}
where $\boldsymbol{\rho} \geq \boldsymbol{0}$ ensures all elements of $\boldsymbol{\rho}$ are non-negative. Using similar steps as in the previous subsection, \eqref{moo-prob-weight-cpa} can be reformulated into a form that can be efficiently solved by standard tools, which is omitted to avoid redundancy.

\section{Weighted-Sum Mismatch Approaches} 
Building on the weighted waveform mismatch minimization approach commonly employed in the \ac{isac} literature to balance sensing and communication performance \cite{fan2018tsp,youli2024twc}, we propose alternative methods for both \ac{fdb} and \ac{cpa} to strike an effective tradeoff between \ac{bp} and \ac{ms} by minimizing the weighted-sum mismatch of two distinct metrics. Specifically, the optimal beamformers, $\overline{\boldsymbol{F}}$ for \ac{bp} and $\underline{\boldsymbol{F}}$ for \ac{ms}, are obtained by solving \eqref{moo-prob-weight-relax} and \eqref{moo-prob-weight-cpa} with $\alpha = 1$ and $\alpha = 0$, respectively. The \emph{balanced} beamformers are then derived from these extremes by applying different strategies: one approach minimizes the weighted-sum mismatch of the beamformers, while the other focuses on minimizing the weighted-sum mismatch of the covariance matrices.

\subsection{Weighted-Sum Mismatch of Beamformers}
\subsubsection{FDB}
Upon obtaining $\overline{\boldsymbol{F}}$ and $\underline{\boldsymbol{F}}$ by solving \eqref{moo-prob-weight-relax} with $\alpha = 1$ and $\alpha = 0$, respectively, we formulate the following optimization problem to minimize the weighted-sum mismatch of beamformers
\begin{subequations}\label{moo-bf-approx}
\begin{align}
\mathop {\min }\limits_{\boldsymbol{F}} \;\;\; 
&\alpha \left\|\boldsymbol{F} - \overline{\boldsymbol{F}} \right\|_{\text{F}}^{2} + \left(1 - \alpha\right)\left\|\boldsymbol{F} - \underline{\boldsymbol{F}} \right\|_{\text{F}}^{2} \label{moo-bf-approx-obj}\\
{\rm{s.t.}}\;\;\; 
&\text{tr}\left(\boldsymbol{F}\boldsymbol{F}^{\mathsf{H}}\right) = P_{\text{B}}/M, \label{moo-bf-approx-power}
\end{align}
\end{subequations}
where \eqref{moo-bf-approx-power} represents the full power transmission constraint. 

Inspired by Remark \ref{remark-opt-struct}, we define $\boldsymbol{F} = \boldsymbol{U}\boldsymbol{\Omega}$, where $\boldsymbol{\Omega}=[\boldsymbol{\omega}_1,\ldots,\boldsymbol{\omega}_{2K+2}] \in \mathbb{C}^{(2K+2)\times (2K+2)}$, ensuring that $\boldsymbol{F}\boldsymbol{F}^{\mathsf{H}}$ satisfies the properties of the optimal covariance matrix. This formulation not only steers the solution toward the desired structure but also decreases complexity by reducing the dimensionality of the variables. Then, \eqref{moo-bf-approx} can be reformulated as
\begin{subequations}\label{moo-bf-approx-ref}
\begin{align}
\mathop {\min }\limits_{\boldsymbol{\Omega}} \;\;\; 
&\left\|\boldsymbol{A}\boldsymbol{U}\boldsymbol{\Omega} - \boldsymbol{B} \right\|_{\text{F}}^{2}  \label{moo-bf-approx-ref-obj}\\
{\rm{s.t.}}\;\;\; 
&\text{tr}\left(\boldsymbol{U}\boldsymbol{\Omega}\boldsymbol{\Omega}^{\mathsf{H}}\boldsymbol{U}^{\mathsf{H}}\right) = P_{\text{B}}/M, \label{moo-bf-approx-ref-power}
\end{align}
\end{subequations}
where $\boldsymbol{A} = [\sqrt{\alpha}\mathbf{I}_{N_{\text{B}}};\sqrt{1-\alpha}\mathbf{I}_{N_{\text{B}}}]$ and $\boldsymbol{B} = [\sqrt{\alpha}\overline{\boldsymbol{F}};\sqrt{1-\alpha}\underline{\boldsymbol{F}}]$. 

The problem remains non-convex due to the equality constraint in \eqref{moo-bf-approx-ref-power}. Notably, we have $\boldsymbol{A}^{\mathsf{H}}\boldsymbol{A} = \mathbf{I}_{N_{\text{B}}}$. By introducing $\boldsymbol{X} = \boldsymbol{U}^{\mathsf{H}}\boldsymbol{A}^{\mathsf{H}}\boldsymbol{B}$ and $\tilde{\boldsymbol{\Omega}}_l = \tilde{\boldsymbol{\omega}}_l\tilde{\boldsymbol{\omega}}_l^{\mathsf{H}}$, where $\tilde{\boldsymbol{\omega}}_l = [1,\boldsymbol{\omega}_l^{\mathsf{T}}]^{\mathsf{T}}$, we can relax the constraint $\text{rank}(\tilde{\boldsymbol{\Omega}}_l) = 1$. This allows us to lift and reformulate \eqref{moo-bf-approx-ref} into 
\begin{subequations}\label{moo-bf-approx-ref-sdp}
\begin{align}
\mathop {\min }\limits_{\tilde{\boldsymbol{\Omega}}_l} \;\;\; 
&\sum_{l=1}^{2K + 2} \text{tr}\left( 
\begin{bmatrix}
0 & -\boldsymbol{X}_{:,l}^{\mathsf{H}}\\
-\boldsymbol{X}_{:,l} & \boldsymbol{U}^{\mathsf{H}}\boldsymbol{U}
\end{bmatrix}
\tilde{\boldsymbol{\Omega}}_l\right)  \label{moo-bf-approx-ref-sdp-obj}\\
{\rm{s.t.}}\;\;\; 
& \sum_{l=1}^{2K + 2} \text{tr}\left(
\begin{bmatrix}
1 & \boldsymbol{0}\\
\boldsymbol{0} & \boldsymbol{U}^{\mathsf{H}}\boldsymbol{U}
\end{bmatrix}
\tilde{\boldsymbol{\Omega}}_l\right) = P_{\text{B}}/M + L, \label{moo-bf-approx-ref-sdp-power} \\
&\left[\tilde{\boldsymbol{\Omega}}_l\right]_{1,1} = 1, \quad \tilde{\boldsymbol{\Omega}}_l \succeq \mathbf{0}, \quad \forall l.
\end{align}
\end{subequations}

The relaxed formulation in \eqref{moo-bf-approx-ref-sdp} results in a convex \ac{sdp} problem. Despite the relaxation, this formulation fits into the class of trust-region subproblems, which are characterized by strong duality and guarantee rank-one solutions \cite{fortin2004trust}. Consequently, the optimal $\tilde{\boldsymbol{\omega}}_l$, and hence the optimal $\boldsymbol{\omega}_l$ for \eqref{moo-bf-approx-ref}, can be recovered from the obtained $\tilde{\boldsymbol{\Omega}}_l$, leading to the determination of $\boldsymbol{F}$.

\subsubsection{CPA}
Let $\boldsymbol{F} = \boldsymbol{U}\boldsymbol{\Upsilon}$, where $\boldsymbol{\Upsilon} = \text{diag}(\boldsymbol{\upsilon})$ and $\boldsymbol{\upsilon}=[\upsilon_1,\ldots,\upsilon_{2K+2}] \in \mathbb{R}^{(2K+2)}$. The matrix $\boldsymbol{\Upsilon}\boldsymbol{\Upsilon}^{\mathsf{H}}$ is diagonal, representing the power allocation. The optimization problem for minimizing the weighted-sum mismatch of beamformers through power allocation is formulated as 
\begin{subequations}\label{moo-bf-approx-cpa}
\begin{align}
\mathop{\min}\limits_{\boldsymbol{\upsilon}} \;\;\; 
&\alpha \left\|\boldsymbol{U}\boldsymbol{\Upsilon} - \overline{\boldsymbol{F}} \right\|_{\text{F}}^{2} + \left(1 - \alpha\right)\left\|\boldsymbol{U}\boldsymbol{\Upsilon}  - \underline{\boldsymbol{F}} \right\|_{\text{F}}^{2} \label{moo-bf-approx-cpa-obj}\\
{\rm{s.t.}}\;\;\; 
&\text{tr}\left(\boldsymbol{U}\boldsymbol{\Upsilon}\boldsymbol{\Upsilon}^{\mathsf{H}} \boldsymbol{U}^{\mathsf{H}}\right) = P_{\text{B}}/M, \label{moo-bf-approx-cpa-power}
\end{align}
\end{subequations}
where $\overline{\boldsymbol{F}}$ and $\underline{\boldsymbol{F}}$ are the beamformers obtained by solving \eqref{moo-prob-weight-cpa} for $\alpha = 1$ and $\alpha = 0$, respectively.

Problem \eqref{moo-bf-approx-cpa} is non-convex. However, after performing algebraic transformations, it can be relaxed as 
\begin{subequations}\label{moo-bf-approx-cpa-sdp}
\begin{align}
\mathop{\min}\limits_{\tilde{\boldsymbol{\Upsilon}}_l} \;\;\; 
&\sum_{l=1}^{2K + 2} \text{tr}\left( 
\begin{bmatrix}
0 & -\Re\left\{\vartheta_l\right\} \\
-\Re\left\{\vartheta_l\right\} & \varpi_l
\end{bmatrix}
\tilde{\boldsymbol{\Upsilon}}_l\right)  \label{moo-bf-approx-cpa-sdp-obj}\\
{\rm{s.t.}}\;\;\; 
& \sum_{l=1}^{2K + 2} \text{tr}\left(
\begin{bmatrix}
1 & 0\\
0 & \boldsymbol{U}_{:,l}^{\mathsf{H}}\boldsymbol{U}_{:,l}
\end{bmatrix}
\tilde{\boldsymbol{\Upsilon}}_l\right) = P_{\text{B}}/M + L, \label{moo-bf-approx-cpa-sdp-power} \\
&\left[\tilde{\boldsymbol{\Upsilon}}_l\right]_{1,1} = 1, \quad \tilde{\boldsymbol{\Upsilon}}_l \succeq \mathbf{0}, \quad \forall l,
\end{align}
\end{subequations}
where $\vartheta_l = \boldsymbol{\Gamma}_{:,l}^{\mathsf{H}}\boldsymbol{B} _{:,l}$ and $\varpi_l = \boldsymbol{\Gamma}_{:,l}^{\mathsf{H}}\boldsymbol{\Gamma}_{:,l}$, with $\boldsymbol{\Gamma} = \boldsymbol{A}\boldsymbol{U}$ and $\tilde{\boldsymbol{\Upsilon}}_l = \tilde{\boldsymbol{\upsilon}}_l\tilde{\boldsymbol{\upsilon}}_l^{\mathsf{H}}$, where $\tilde{\boldsymbol{\upsilon}}_l = [1,\upsilon_l]^{\mathsf{T}}$. Similarly, by invoking strong duality, the rank-one solutions are guaranteed\cite{fortin2004trust}. The optimal $\tilde{\boldsymbol{\upsilon}}_l$, and consequently the optimal $\boldsymbol{\upsilon}$ for \eqref{moo-bf-approx-cpa}, can be derived from the obtained $\tilde{\boldsymbol{\Upsilon}}_l$, leading to obtaining $\boldsymbol{F}$.

\subsection{Weighted-Sum Mismatch of Covariance Matrices}
\subsubsection{FDB}
We observe that the position-domain \ac{fim}, and thus the \acp{crb} in \eqref{crb-pos-bi} and \eqref{crb-pos-mono}, are directly influenced by $\boldsymbol{F}\boldsymbol{F}^{\mathsf{H}}$. This term can be interpreted as the \emph{covariance matrix} of the transmit signal (up to scaling), capturing the combined effect of different beamformers. Motivated by this insight, we propose minimizing the weighted sum mismatch of these covariance matrices. From the resulting covariance matrix, the beamformers can then be derived. Specifically, using the optimal form of the covariance matrix, represented as $\boldsymbol{V} = \boldsymbol{U}\boldsymbol{\Lambda}\boldsymbol{U}^{\mathsf{H}}$, we formulate the optimization problem as 
\begin{subequations}\label{moo-bf-approx-var}
\begin{align}
\mathop {\min }\limits_{\boldsymbol{\Lambda}} \;\;\; 
&\alpha \left\|\boldsymbol{U}\boldsymbol{\Lambda}\boldsymbol{U}^{\mathsf{H}} - \overline{\boldsymbol{V}} \right\|_{\text{F}}^{2} + \left(1 - \alpha\right)\left\|\boldsymbol{U}\boldsymbol{\Lambda}\boldsymbol{U}^{\mathsf{H}} - \underline{\boldsymbol{V}} \right\|_{\text{F}}^{2} \label{moo-bf-approx-var-obj}\\
{\rm{s.t.}}\;\;\; 
&\text{tr}\left(\boldsymbol{U}\boldsymbol{\Lambda}\boldsymbol{U}^{\mathsf{H}}\right) = P_{\text{B}}/M, \label{moo-bf-approx-var-power}\\
&\boldsymbol{\Lambda} \succeq \boldsymbol{0}, \label{moo-bf-approx-var-rank}
\end{align}
\end{subequations}
where $\overline{\boldsymbol{V}} = \overline{\boldsymbol{F}}\overline{\boldsymbol{F}}^{\mathsf{H}}$ and $\underline{\boldsymbol{V}} = \underline{\boldsymbol{F}}\underline{\boldsymbol{F}}^{\mathsf{H}}$.

The problem in \eqref{moo-bf-approx-var} is a \ac{sdp}, which can be efficiently solved using tools like CVX. Once the solution is obtained, the beamformers $\boldsymbol{F}$ can be extracted through matrix decomposition techniques.

\subsubsection{CPA}
For the weighted-sum mismatch of covariance matrices approach under the \ac{cpa} scheme, we replace $\boldsymbol{\Lambda}$ in \eqref{moo-bf-approx-var} with $\boldsymbol{U}\text{diag}\left(\boldsymbol{\rho}\right)\boldsymbol{U}^{\mathsf{H}}$ and modify \eqref{moo-bf-approx-var-rank} to $\boldsymbol{\rho} \geq \boldsymbol{0}$. The resulting formulation is a convex \ac{qcqp} problem that can be solved using standard tools like CVX. For brevity, detailed steps are omitted.

\begin{remark}\label{reamrk-guiding-solution}
We would like to highlight that while there are multiple methods to decompose the covariance matrix $\boldsymbol{V}$ into beamformers $\boldsymbol{F}$ such that $\boldsymbol{V} = \boldsymbol{F}\boldsymbol{F}^{\mathsf{H}}$, in the proposed weighted-sum mismatch of beamformers approach, the guiding beamformers $\overline{\boldsymbol{F}}$ and $\underline{\boldsymbol{F}}$ under the \ac{fdb} (\ac{cpa}) scheme must be expressed as $\boldsymbol{U}\boldsymbol{\Omega}$ ($\boldsymbol{U}\boldsymbol{\Upsilon}$). This ensures that the guiding beamformers are aligned with the construction of the balanced beamformers, thereby reducing the risk of unnecessary mismatches. On the other hand, in the weighted-sum mismatch of covariance matrices approach, the guiding solutions are presented as variance matrices in the \ac{bp} and \ac{ms} scenarios, which are represented in a fixed form, unlike beamformers that can take various forms.
\end{remark}

\section{Analog beamforming approaches}
The previous section introduced digitally designed beamformers, allowing individual control over both the amplitude and phase of each element within the beamformer. While this digital approach offers greater design flexibility, it also incurs higher hardware complexity and costs, as each antenna element requires a dedicated, high-cost \ac{dac}. As communication networks advance towards higher frequency bands and larger antenna arrays, these hardware demands may become increasingly difficult to manage. In this section, we propose analog beamforming methods, where all elements in a beamformer share the same amplitude but allow for individual phase adjustments through more economical \acp{ps}, applicable under both the \ac{fdb} and \ac{cpa} schemes. This approach can significantly reduce hardware costs by enabling multiple antennas to share a common \ac{dac}, providing a cost-effective complement to the proposed digital beamforming methods. In developing the analog beamforming approach, we adopt the weighted-sum mismatch framework discussed in the previous section for simplicity, rather than the weighted-sum \ac{crb} optimization paradigm. Additionally, as shown in the numerical results, minimizing the mismatch of covariance matrices proves to be more effective than minimizing the mismatch of beamformers. Consequently, our analog design prioritizes the former approach.

\subsection{Full-Dimensional Beamforming}
To balance the tradeoff between \ac{bp} and \ac{ms} under the analog \ac{fdb} design, the mismatch-minimization optimization problem is formulated as 
\begin{subequations}\label{moo-analog-bf-fdb}
\begin{align}
\mathop {\min }\limits_{\rho_l,\phi_{i,l}} \;\;\; 
&\alpha \left\|\boldsymbol{F}\boldsymbol{F}^{\mathsf{H}} - \overline{\boldsymbol{V}} \right\|_{\text{F}}^{2} + \left(1 - \alpha\right)\left\|\boldsymbol{F}\boldsymbol{F}^{\mathsf{H}} - \underline{\boldsymbol{V}} \right\|_{\text{F}}^{2} \label{moo-analog-bf-fdb-obj}\\
{\rm{s.t.}}\;\;\; 
&\phi_{i,l}\in \left[0,2\pi\right], \quad \forall i,l, \label{moo-analog-bf-fdb-constr}  \\
&\sum_{l=1}^{L} \rho_l = \frac{P}{M}, \quad \boldsymbol{\rho} \ge \boldsymbol{0},  \label{moo-analog-bf-fdb-power}
\end{align}
\end{subequations}
where $[\boldsymbol{F}]_{i,l} = \sqrt{\rho_l/N_{\text{B}}}e^{\jmath \phi_{i,l}}$ and constraint \eqref{moo-analog-bf-fdb-constr} ensures the analog nature of the beamformers, meaning that all elements of a given beamformer have equal amplitude, with phases controlled by \acp{ps}.

This problem is non-convex, making it challenging to solve directly. To address this, we propose an \ac{ao} framework that tackles \eqref{moo-analog-bf-fdb} iteratively. We first reformulate \eqref{moo-analog-bf-fdb} into an equivalent problem as
\begin{subequations}\label{moo-analog-bf-fdb-reca}
\begin{align}
\mathop {\min }\limits_{\rho_l,\phi_{i,l}} \;\;\; 
& \text{tr}\left(\boldsymbol{F}^{\mathsf{H}}\boldsymbol{F}\boldsymbol{F}^{\mathsf{H}}\boldsymbol{F}\right) - 2 \Re \left\{\text{tr}\left(\boldsymbol{F}^{\mathsf{H}} \boldsymbol{\Psi} \boldsymbol{F} \right) \right\} \label{moo-analog-bf-fdb-reca-obj}\\
{\rm{s.t.}}\;\;\; 
&\phi_{i,l}\in \left[0,2\pi\right], \quad \forall i,l, \label{moo-analog-bf-fdb-reca-constr}  \\
&\sum_{l=1}^{L} \rho_l = \frac{P}{M}, \quad \boldsymbol{\rho} \ge \boldsymbol{0},  \label{moo-analog-bf-fdb-reca-power}
\end{align}
\end{subequations}
where $\boldsymbol{\Psi} = \boldsymbol{A}^{\mathsf{H}}\boldsymbol{Q}$, and $\boldsymbol{Q} = [\sqrt{\alpha}\overline{\boldsymbol{V}};\sqrt{1-\alpha}\underline{\boldsymbol{V}}]$. Then, by substituting $[\boldsymbol{F}]_{i,l} = \sqrt{\rho_l/N_{\text{B}}}e^{\jmath \phi_{i,l}}$ into \eqref{moo-analog-bf-fdb-reca-obj}, we expand the objective function into $f(\boldsymbol{\rho},\boldsymbol{\Phi}) = \sum_{i=1}^{L}\sum_{k=1}^{L}\sum_{n=1}^{N_{\text{B}}}\sum_{q=1}^{N_{\text{B}}}\rho_i \rho_k \cos (\phi_{n,k} - \phi_{n,i} + \phi_{q,i} - \phi_{q,k}) - 2N_{\text{B}} \sum_{i=1}^{L}\sum_{n=1}^{N_{\text{B}}}\sum_{q=1}^{N_{\text{B}}} \rho_i (\Re \{\boldsymbol{\Psi}_{n,q} \} \cos (\phi_{q,i} - \phi_{n,i}) - \Im \{\boldsymbol{\Psi}_{n,q}\} \sin (\phi_{q,i} - \phi_{n,i}) ) $,
where the matrix $\boldsymbol{\Phi} \in \mathbb{R}^{N_{\text{B}} \times L}$ contains all the phases across different beamformers.

\subsubsection{Power Allocation Subproblem}  
Given fixed values of $\boldsymbol{\Phi}$, the optimization problem over $\boldsymbol{\rho}$ can be expressed as 
\begin{subequations}\label{moo-analog-bf-fdb-pow-alloc}  
\begin{align}  
\mathop {\min }\limits_{\boldsymbol{\rho}} \;\;\;  
& f\left(\boldsymbol{\rho},\boldsymbol{\Phi}\right) \label{moo-analog-bf-fdb-pow-alloc-obj}\\  
{\rm{s.t.}}\;\;\;  
&\sum_{l=1}^{L} \rho_l = P_{\text{B}}/M, \quad \boldsymbol{\rho} \geq \boldsymbol{0}.  
\end{align}  
\end{subequations}  

The objective function presents a high degree of nonlinearity, which we address by utilizing the \ac{sqp} framework. This method iteratively solves a sequence of quadratic subproblems that approximate the original smooth, nonlinear optimization problem with both inequality and equality constraints \cite{wright2006opt}. To aid comprehension, Appendix \ref{sqp-fundamental} outlines the key concepts and basic steps of the \ac{sqp} framework. However, the detailed derivations of terms in \eqref{moo-analog-bf-fdb-pow-alloc} are omitted for brevity. Readers seeking the complete procedure are directed to Algorithm 18.3 in \cite{wright2006opt}.

\subsubsection{Phase Optimization Subproblem}  
The optimization problem with respect to $\boldsymbol{\Phi}$, given fixed values of $\boldsymbol{\rho}$, is formulated as follows:  
\begin{subequations}\label{moo-analog-bf-fdb-phase-opt}  
\begin{align}  
\mathop {\min }\limits_{\boldsymbol{\Phi}} \;\;\;  
& f\left(\boldsymbol{\rho},\boldsymbol{\Phi}\right) \label{moo-analog-bf-fdb-phase-opt-obj}\\  
{\rm{s.t.}}\;\;\;  
& \phi_{i,l} \in \left[0, 2\pi \right], \quad \forall i,l.  
\end{align}  
\end{subequations}  

This subproblem can similarly be solved using the \ac{sqp} framework. Algorithm \ref{sqp-anlog-bf} outlines the detailed procedure for designing the analog \ac{fdb} with \ac{ao} framework.

\renewcommand{\algorithmicrequire}{\textbf{Input:}}
\renewcommand{\algorithmicensure}{\textbf{Output:}}
\begin{algorithm}[t]
\caption{Algorithm for Solving \eqref{moo-analog-bf-fdb}}
\label{sqp-anlog-bf}       %
\begin{algorithmic}[1]
% \State \textbf{Input}: {$\overline{\boldsymbol{V}}$, $\underline{\boldsymbol{V}}$, $\alpha$;}
\State \textbf{Initialize}: {$\boldsymbol{\Phi}$;}
\Repeat
\State {Update $\boldsymbol{\rho}$ by solving \eqref{moo-analog-bf-fdb-pow-alloc} using the \ac{sqp} framework;}
\State {Update $\boldsymbol{\Phi}$ by solving \eqref{moo-analog-bf-fdb-phase-opt} using the \ac{sqp} framework;}
\Until {the reduction ratio of the objective value falls below a specified threshold;}
\State {Recover $\boldsymbol{F}$ from $\boldsymbol{\rho}$ and $\boldsymbol{\Phi}$;}
\State \textbf{Output}: {$\boldsymbol{F}$.}
\end{algorithmic}
\end{algorithm}

\subsection{Codebook-based Power Allocation}
To achieve analog \ac{cpa}, a crucial step is selecting an effective analog codebook, which ensures meaningful power allocation. Inspired by the digital \ac{cpa} framework, we construct the analog codebook directly from the \emph{optimal} digital codebook matrix $\boldsymbol{U}$, rather than designing it from scratch. The main idea is to retain the first $K+1$ codewords (columns) of $\boldsymbol{U}$, as they are already in analog form, and to approximate the last $K+1$ codewords with analog counterparts that exhibit similar beampatterns. This approach minimizes the performance gap between power allocation over the original digital codebook and the newly constructed analog codebook, thereby preserving the achievable performance as much as possible. Specifically, mimicking $\dot{\boldsymbol{a}}_{\text{B}}\left(\theta_l\right)$ with the analog codeword $\tilde{\boldsymbol{f}}_l$ under the principle of beampattern approximation is formulated as an optimization problem, given by
\begin{subequations}\label{analog-code-construct}  
\begin{align}  
\mathop {\min }\limits_{A_l,\phi_{i,l}} \;\;\;  
& \left\|\boldsymbol{T}\tilde{\boldsymbol{f}}_l - \boldsymbol{T}\dot{\boldsymbol{a}}_{\text{B}}\left(\theta_l\right)\right\|^2 \label{manalog-code-construct-obj}\\  
{\rm{s.t.}}\;\;\;  
&\phi_{i,l} \in \left[0, 2\pi \right], \quad \forall i,l,
\end{align}  
\end{subequations}  
where $\left[\tilde{\boldsymbol{f}}_l\right]_i = A_l e^{\jmath \phi_{i,l}}$ and $A_l$ is the amplitude to be determined, and $\boldsymbol{T}=[\boldsymbol{a}^{\mathsf{H}}_{\text{B}}(0);\boldsymbol{a}^{\mathsf{H}}_{\text{B}}(\pi/\tilde{N});\ldots;\boldsymbol{a}^{\mathsf{H}}_{\text{B}}((\tilde{N}-1)\pi/\tilde{N})]$ is a matrix containing transmit steering vectors covering a complete angular period of $\pi$, with $\tilde{N}$ being the number of candidate angles.

% We observe that the amplitude and phase terms are coupled in the objective function, which makes solving \eqref{analog-code-construct} directly challenging. To address this, we apply the \ac{ao} framework, allowing us to iteratively optimize the amplitude and phase components separately\footnote{For notational simplicity, the subscript $l$ is omitted in the subsequent derivations.}.

% \subsubsection{Amplitude Optimization Subproblem}
% Fixing $\phi_{i}$, the optimization problem for $A$ can be reformulated as:
% \begin{equation}\label{analog-code-construct-amp}  
% \mathop {\min }\limits_{A} \;\;\;  
% \left\| A \boldsymbol{T} \tilde{\boldsymbol{v}} - \boldsymbol{T}\dot{\boldsymbol{a}}_{\text{B}}\left(\theta\right)  \right\|^2,
% \end{equation}  
% where $\tilde{\boldsymbol{v}} = [e^{\jmath \phi_{1}},\ldots,e^{\jmath \phi_{N_{\text{B}}}}]^{\mathsf{T}}$. The optimal solution to \eqref{analog-code-construct-amp} is given by
% \begin{equation}\label{analog-opt-amp}
% A = \frac{\Re \left\{\dot{\boldsymbol{a}}_{\text{B}}^{\mathsf{H}}\left(\theta\right)\boldsymbol{T}^{\mathsf{H}}\boldsymbol{T} \tilde{\boldsymbol{v}}\right\}}{\tilde{\boldsymbol{v}}^{\mathsf{H}} \boldsymbol{T}^{\mathsf{H}} \boldsymbol{T} \tilde{\boldsymbol{v}}}.
% \end{equation}

% \usepackage{multirow}
\begin{table*}[t]
\caption{complexity of proposed approaches}
\vspace{-4mm}
\centering
\label{complexity}
\begin{center}
\scalebox{1}{
\begin{tabular}{|c|ccc|cc|}
\hline
\multirow{2}{*}{\textbf{Taxonomy}} &
  \multicolumn{3}{c|}{\textbf{Digital}} &
  \multicolumn{2}{c|}{\textbf{Analog}} \\ \cline{2-6} 
 &
  \multicolumn{1}{c|}{WCRB} &
  \multicolumn{1}{c|}{WBF} &
  WCM &
  \multicolumn{1}{c|}{Power optimization} &
  Phase optimization \\ \hline
FDB &
  \multicolumn{1}{c|}{$\mathcal{O}\left(K^6\right)$} &
  \multicolumn{1}{c|}{$\mathcal{O}\left(K^8\right)$} &
  $\mathcal{O}\left(K^6\right)$ &
  \multicolumn{1}{c|}{$\mathcal{O}\left(L^3\right)$} &
  $\mathcal{O}\left(N_{\text{B}}^3 L^3\right)$ \\ \hline
CPA &
  \multicolumn{1}{c|}{$\mathcal{O}\left(K^6\right)$} &
  \multicolumn{1}{c|}{$\mathcal{O}\left(K^4\right)$} &
  $\mathcal{O}\left(K^3\right)$ &
  \multicolumn{2}{c|}
  {$\mathcal{O}\left(N_{\text{B}}\right)$} \\ \hline
\end{tabular}
}
\end{center}
\end{table*}

Problem \eqref{analog-code-construct} can be solved using gradient projection \cite{tranter2017tsp}, through which the analog codeword $\tilde{\boldsymbol{f}}_l$  corresponding to $\dot{\boldsymbol{a}}_{\text{B}}\left(\theta_l\right)$can be generated. The analog codebook is defined as $\tilde{\boldsymbol{U}} = [\boldsymbol{a}_{\text{B}}\left(\theta_0\right), \ldots, \boldsymbol{a}_{\text{B}}\left(\theta_K\right), \tilde{\boldsymbol{f}}_1, \ldots, \tilde{\boldsymbol{f}}_{K+1}]$. The remaining steps of the analog \ac{cpa} are essentially the same as those for digital \ac{cpa} introduced in Section IV-B and are omitted here for brevity.

\section{Convergence and Complexity Analysis}
\subsection{Convergence}
The convergence of the digital beamforming approaches presented in Sections III and IV is straightforward, as they involve solving a one-shot convex optimization problem. For the analog beamforming approaches discussed in Section V, developed under the \ac{ao} framework, convergence is expected because each subproblem converges to a stationary point, with the objective values being bounded and non-increasing. Specifically, for the analog \ac{fdb} approach, convergence is ensured by the \ac{sqp} framework's convergence properties \cite{wright2006opt}. For the analog \ac{cpa} approach, convergence is evident by recognizing the convergence of the analog codeword method achieved through gradient projection \cite{tranter2017tsp}.

\subsection{Complexity Analysis}
\subsubsection{Digital Schemes}
The per-iteration computational complexity of solving an \ac{sdp} problem using the interior-point method is given by $\mathcal{O}(I^2 \sum_{j=1}^{J}d_j^2 + I \sum_{j=1}^{J}d_j^3)$, where $I$ and $J$ represent the numbers of optimization variables and \ac{lmi} constraints, respectively, and $d_j$ denotes the row/column dimension of the matrix associated with the $j$-th \ac{lmi} constraint\cite{furkan2022tvt}. In comparison, the per-iteration complexity of solving a \ac{qcqp} problem is approximately $\mathcal{O}(d^3)$, where $d$ represents the number of optimization variables\cite{boyd}. The dimensional parameters for the proposed digital\footnote{For brevity, the prefix ``Digital'' is omitted from the following abbreviations unless specified for emphasis.} beamforming schemes are as follows:
\begin{itemize}
    \item \ac{fdb}-Weighted-Sum \ac{crb} (WCRB): The \ac{sdp} problem associated with \eqref{moo-prob-weight-relax}, where $I = (2K+2)^2 + (2K+2)^2 + 4$, $J = 5$, $d_1 = 4K+6$, $d_2 = 4K+4$, $d_3 = 2$, $d_4 = 2K+2$, and $d_5 = 2K+2$.
    \item \ac{fdb}-Weighted-Sum Beamformer Mismatch (WBF): The \ac{sdp} problem associated with \eqref{moo-bf-approx-ref-sdp}, where $I = (2K+2)(2K+3)^2$, $J = 2K+2$, and $d_j = 2K+3$.
    \item \ac{fdb}-Weighted-Sum Covariance Matrix Mismatch (WCM): The \ac{sdp} problem associated with \eqref{moo-bf-approx-var}, where $I = (2K+2)^2$, $J = 1$, and $d_1 = 2K+2$.
    \item \ac{cpa}-WCRB: The \ac{sdp} problem associated with \eqref{moo-prob-weight-cpa}, where $I = (2K+2)^2 + (2K+2) + 4$, $J = 4$, $d_1 = 4K+6$, $d_2 = 4K+4$, $d_3 = 2$, and $d_4 = 2K+2$.
    \item \ac{cpa}-WBF: The \ac{sdp} problem associated with \eqref{moo-bf-approx-cpa-sdp}, where $I = 4(2K+2)$, $J = 2K+2$, and $d_j = 2$.
    \item \ac{cpa}-WCM: The \ac{qcqp} problem described in Section-IV-B-2, where $d = 2K+2$.
\end{itemize} 

\begin{figure}[t]
\centering 
\centerline{% This file was created by matlab2tikz.
\definecolor{mycolor1}{rgb}{0.00000,0.44700,0.74100}%
\begin{tikzpicture}

\begin{axis}[%
width=72mm,
height=40mm,
at={(0mm, 0mm)},
scale only axis,
xmin=0,
xmax=6,
xticklabel style = {font=\color{white!15!black},font=\footnotesize},
xlabel style={font=\footnotesize, xshift=2mm},
xlabel={Iteration},
ymode=log,
ymin=0.481947414378764,
ymax=0.724417491136977,
yminorticks=true,
yticklabel style = {font=\color{white!15!black},font=\footnotesize},
ylabel style={font=\footnotesize, yshift=0mm,font=\footnotesize},
ylabel={Objective value},
axis background/.style={fill=white},
xmajorgrids,
ymajorgrids,
yminorgrids,
legend style={font=\scriptsize, legend cell align=left, align=left, draw=white!15!black}
]
\addplot [color=mycolor1, line width=2.0pt]
  table[row sep=crcr]{%
0	0.724417491136977\\
1	0.724417491136977\\
2	0.675192367158341\\
3	0.532144066126704\\
4	0.492744619875574\\
5	0.482453490000415\\
6	0.481947414378764\\
};
\addlegendentry{Analog \ac{fdb} via Algorithm \ref{sqp-anlog-bf}}

\end{axis}
\end{tikzpicture}%
}
\caption{Illustrations of the convergence behavior for solving the analog \ac{fdb} problem using Algorithm \ref{sqp-anlog-bf}.}
% \vspace{-5mm}
\label{converence-plot}
\end{figure}
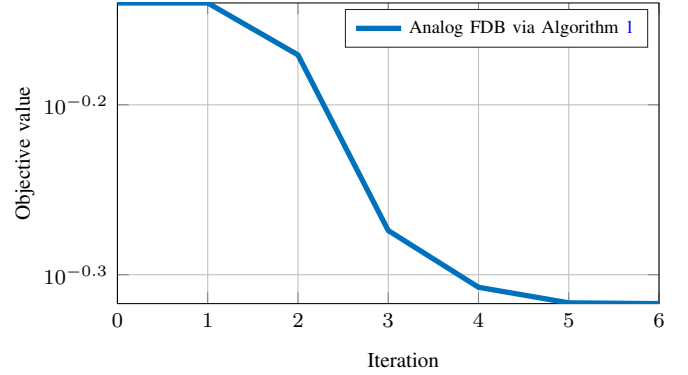

\begin{figure}[t]
\centering 
\centerline{\input{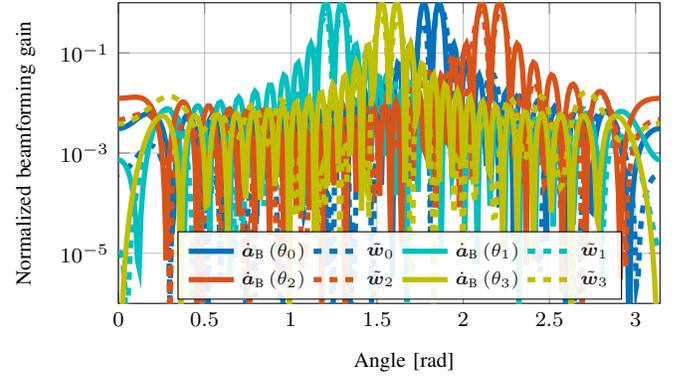}}
\caption{Comparison of beampatterns between derivative codewords and their analog counterparts.}
% \vspace{-5mm}
\label{analog-codebook-compare}
\end{figure}

\begin{figure*}[t]
\begin{minipage}[b]{0.99\linewidth}
\begin{center}
\end{center}
\centering
\centerline{\includegraphics[width=1\linewidth]{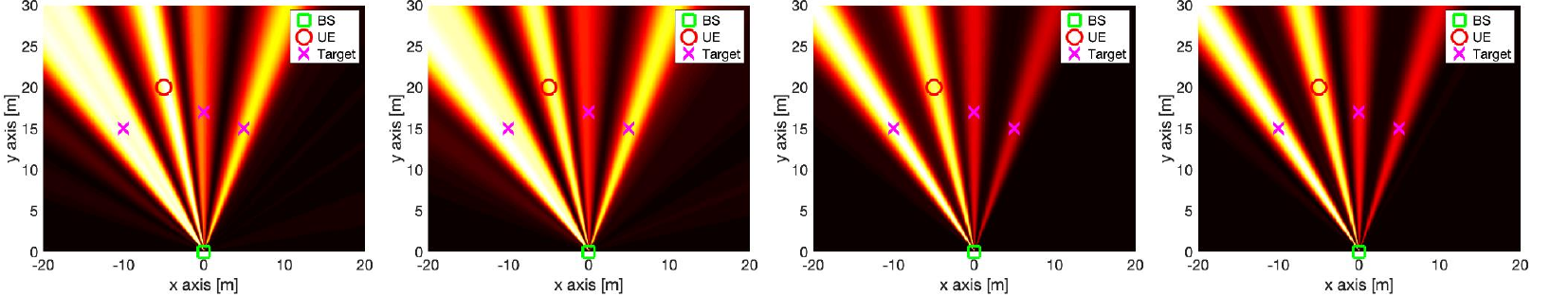} }
% \include{Figures/fig-1-1}
% \vspace{-0.8cm}
\small
\centerline{
(a) \ac{bp}: Digital-FDB 
\hspace{1.5cm} 
(b) \ac{bp}: Analog-FDB 
\hspace{1.5cm} 
(c) \ac{bp}: Digital-CPA
\hspace{1.5cm} 
(d) \ac{bp}: Analog-CPA
}
\vspace{0.15cm}
% \centerline{(a) Scenario 1: Two RISs are parallel facing each other.}
\normalsize
\end{minipage}

\begin{minipage}[b]{0.99\linewidth}
\centering
\centerline{\includegraphics[width=1\linewidth]{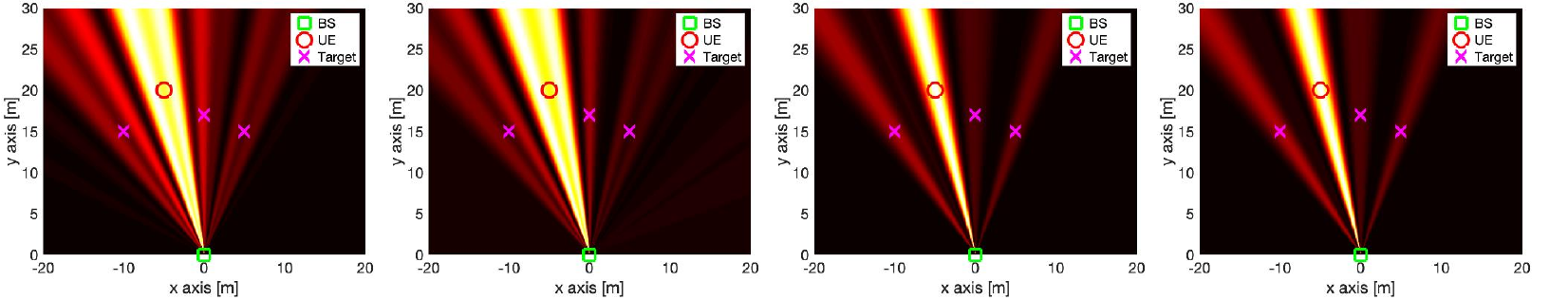}}
\small
\centerline{
(e) \ac{ms}: Digital-FDB 
\hspace{1.5cm} 
(f) \ac{ms}: Analog-FDB 
\hspace{1.5cm} 
(g) \ac{ms}: Digital-CPA
\hspace{1.5cm} 
(h) \ac{ms}: Analog-CPA
}
% \centerline{(b) Scenario 2: Two RISs are located on the same plane facing the +x axis.}
\normalsize
\end{minipage}

\caption{2D beampatterns generated by different schemes, where the first row shows results for \ac{bp} and the second row for \ac{ms}: (a) Digital-FDB; (b) Analog-FDB; (c) Digital-CPA; (d) Analog-CPA; (e) Digital-FDB; (f) Analog-FDB; (g) Digital-CPA; and (h) Analog-CPA.}
\vspace{-3mm}
\label{bp}
\end{figure*}

\subsubsection{Analog Schemes}
The complexity of the proposed \ac{ao}-based analog beamforming schemes is determined by both the number of iterations and the complexity of each iteration. The detailed analysis is as follows:

\begin{itemize}
    \item Analog-\ac{fdb}: This scheme is addressed through \ac{ao}, alternating between two \ac{sqp} problems, each leading to inner quadratic subproblems classified as \ac{qcqp} problems. The total complexity is expressed as $I_{\text{Ana}}^{\text{FDB-Pow}}C_{\text{Ana}}^{\text{FDB-Pow}} + I_{\text{Ana}}^{\text{FDB-Pha}}C_{\text{Ana}}^{\text{FDB-Pha}}$, where $I_{\text{Ana}}^{\text{FDB-Pow}}$ and $I_{\text{Ana}}^{\text{FDB-Pha}}$ represent the total iterations for the quadratic subproblems related to power allocation and phase optimization, respectively. The terms $C_{\text{Ana}}^{\text{FDB-Pow}}$ and $C_{\text{Ana}}^{\text{FDB-Pha}}$ refer to the complexity of solving these subproblems, respectively.
    
    \item Analog-\ac{cpa}: The primary source of complexity in this scheme stems from the analog codebook construction, so the complexity of power allocation optimization after the codebook is obtained is not considered. The complexity for constructing $K+1$ codewords is given by $(K + 1)I_{\text{Ana}}^{\text{CPA}}C_{\text{Ana}}^{\text{CPA}}$, where $I_{\text{Ana}}^{\text{CPA}}$ represents the total iterations for the inner closed-form updates of the adopted gradient projection method\cite{tranter2017tsp} when constructing a specific codeword. The terms $C_{\text{Ana}}^{\text{CPA}}$ denotes the complexity of computing the updates.
\end{itemize}

For clarity, the dominant per-iteration complexity of the proposed beamforming schemes is derived, simplified, and summarized in Table \ref{complexity}. It is worth emphasizing that in high-frequency scenarios, such as millimeter-wave or even terahertz systems, the channel typically exhibits sparsity, leading to $K \ll N_{\text{B}}$. This sparsity allows us to exploit the optimal covariance matrix structure, thereby significantly reducing the computational complexity of the proposed digital beamforming approaches, which is now governed by an order of $K$.

\section{Numerical Results}
\subsection{Scenarios}
%%%%%%%%%%%%%%%%%%%%%%%%% Simulation parameters %%%%%%%%%%%%%%%%%%%%%%%%%%
Unless specified otherwise, the simulation parameters are configured as follows: The \ac{bs} is equipped with $N_{\text{B}} = 16$ transmit/receive antennas, located at $\boldsymbol{p}_{\text{B}} = [0\text{ m}, 0\text{ m}]^{\mathsf{T}}$. The \ac{ue}, with $N_\text{U} = 16$ antennas, is positioned at $\boldsymbol{p}_{\text{U}} = [-5\text{ m}, 20\text{ m}]^{\mathsf{T}}$. There are $K=3$ targets, placed at $\boldsymbol{p}_1 = [-10\text{ m}, 15\text{ m}]^{\mathsf{T}}$, $\boldsymbol{p}_2 = [5\text{ m}, 15\text{ m}]^{\mathsf{T}}$, and $\boldsymbol{p}_3 = [0\text{ m}, 17\text{ m}]^{\mathsf{T}}$.
The transmit power is set at $P_{\text{B}} = -20\text{ dBm}$, with a carrier frequency of $f_c = 28 \text{ GHz}$ and a bandwidth of $W = 120 \text{ MHz}$. The system uses $M = 1024$ subcarriers, with a noise figure of $F = 10 \text{ dB}$ and noise \ac{psd} $N_0 = -173.855 \text{ dBm/Hz}$. The simulation includes $L = 16$ slots, each comprising $P = 100$ pilot symbols, with a clock bias $\Delta t = 1\;\mu\text{s}$ and a relative \ac{ue} orientation of $\Delta \phi = (110/180)\pi$.
{The channel gains are generated using a standard free-space path loss model\cite{henk2022cl}. For the $k$-th path, the phases $\overline{\zeta}_k$ (for \ac{bp}) and $\underline{\zeta}_k$ (for \ac{ms}) are uniformly distributed over $[-\pi, \pi]$. In \ac{bp}, the \ac{los} channel gain is given by $\overline{\beta}_0 = e^{\jmath \overline{\zeta}_0} \lambda/(4 \pi \left\|\boldsymbol{p}_{\text{B}} - \boldsymbol{p}_{\text{U}}\right\|)$, while the \ac{nlos} channel gain is expressed as $\overline{\beta}_k = \overline{\sigma}_{\text{RCS},k}e^{\jmath \overline{\zeta}_k} \lambda/((4 \pi)^{3/2} \left\|\boldsymbol{p}_{\text{U}} - \boldsymbol{p}_k\right\| \left\|\boldsymbol{p}_k - \boldsymbol{p}_{\text{B}}\right\|)$. For \ac{ms}, the channel gain is expressed as $\underline{\beta}_k = \underline{\sigma}_{\text{RCS},k}e^{\jmath \overline{\zeta}_k} \lambda/((4 \pi)^{3/2} \left\|\boldsymbol{p}_k - \boldsymbol{p}_{\text{B}}\right\|^2)$. Here, $\overline{\sigma}_{\text{RCS},k}$ (for \ac{bp}) and $\underline{\sigma}_{\text{RCS},k}$ (for \ac{ms}) represent the \ac{rcs} of the $k$-th target. Specifically, $\underline{\sigma}_{\text{RCS},0} = 10 \text{ m}^2$, while $\overline{\sigma}_{\text{RCS},k} = \underline{\sigma}_{\text{RCS},k} = 100 \text{ m}^2$ ($k=1,\ldots,K$)\footnote{In \ac{ms}, the \ac{rcs} of the \ac{ue} is set lower than that of other passive objects because the \ac{ue} (such as a handset) is typically much smaller than environmental objects like rocks or vehicles, which have higher \acp{rcs} due to their size and structure.}. The wavelength $\lambda$ is defined as $\lambda = c/f_c$, where $c$ is the speed of light.
%%%%%%%%%%%%%%%%%%%%%%%%% Baseline Explaination %%%%%%%%%%%%%%%%%%%%%%%%%%
\subsection{Compared Schemes}
We evaluate system performance using the proposed digital and analog beamforming approaches. Specifically, for the digital methods, we investigate the six schemes outlined in Section VI-B-1: \ac{fdb}-WCRB, \ac{fdb}-WBF, \ac{fdb}-WCM, \ac{cpa}-WCRB, \ac{cpa}-WBF, and \ac{cpa}-WCM. Additionally, we examine the fused \ac{bp}-\ac{ms} case described in Remark \ref{remark-fusion} within the digital-\ac{fdb}-WCRB paradigm, which we refer to simply as ``Fusion''. For the analog methods, we evaluate the two schemes outlined in Section VI-B-2: Analog-\ac{fdb} and Analog-\ac{cpa}.

% \subsubsection{Digital} The first category focuses on solving \ac{fdb} optimization problems, represented by \eqref{moo-prob-weight}, \eqref{moo-bf-approx}, and \eqref{moo-bf-approx-var}. These are further categorized as \ac{fdb}-weighted \ac{crb} (\ac{fdb}-WCRB), \ac{fdb}-weighted beamformer (\ac{fdb}-WBF), and \ac{fdb}-weighted covariance matrix (\ac{fdb}-WCM), respectively.
% As a low-complexity alternative, we introduce a \ac{cpa} approach. The core idea is that 
% These methods are further categorized as \ac{cpa} weighted \ac{crb} (\ac{cpa}-WCRB), \ac{cpa} weighted beamformer (\ac{cpa}-WBF), and \ac{cpa} weighted covariance matrix (\ac{cpa}-WCM), respectively. Lastly, a simple baseline is the \ac{apa} across the given codebook $\boldsymbol{U}$, where no distinction or tradeoff is made between \ac{bp} and \ac{ms}. 

% \subsubsection{Analog}    

% \subsubsection{Fusion}    
%%%%%%%%%%%%%%%%%%%%%%%%%%%%%%%%%%%%%%%%%%%%%%%%%%%%%%%%%%%%%%%%%%%%%%%%%%

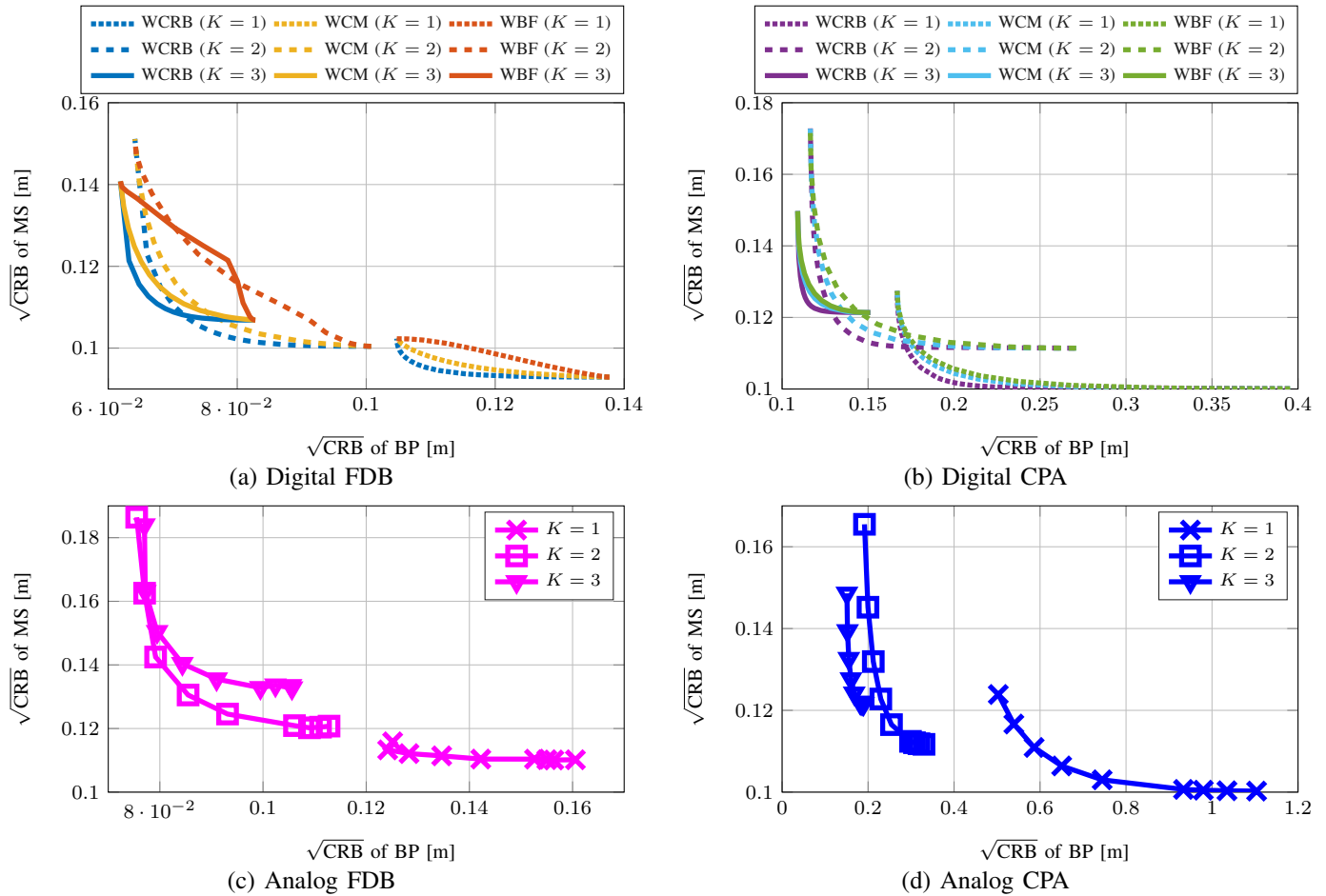
\begin{figure*}[t]
\centering
\begin{minipage}[b]{0.48\linewidth}
  \centering
\definecolor{mycolor1}{rgb}{0.00000,0.44700,0.74100}%
\definecolor{mycolor2}{rgb}{0.92900,0.69400,0.12500}%
\definecolor{mycolor3}{rgb}{0.85000,0.32500,0.09800}%
\begin{tikzpicture}

\begin{axis}[%
width=72mm,
height=40mm,
at={(0mm, 0mm)},
scale only axis,
xmin=0.06,
xmax=0.14,
xticklabel style = {font=\color{white!15!black},font=\footnotesize},
xlabel style={font=\footnotesize, xshift=2mm},
xlabel={$\sqrt{\text{CRB}}$ of \ac{bp} [m]},
ymin=0.09,
ymax=0.16,
yminorticks=true,
yticklabel style = {font=\color{white!15!black},font=\footnotesize},
ylabel style={font=\footnotesize, yshift=0mm,font=\footnotesize},
ylabel={$\sqrt{\text{CRB}}$ of \ac{ms} [m]},
axis background/.style={fill=white},
xmajorgrids,
ymajorgrids,
yminorgrids,
legend style={font=\scriptsize, at={(1.0,1.34)}, anchor=north east, legend cell align=left, align=left, fill = white, fill opacity=0.9, legend columns = 3}
]
\addplot [color=mycolor1, densely dotted, line width=2.0pt]
  table[row sep=crcr]{%
0.137701725531466	0.0929151562378486\\
0.130024050974744	0.0929608916504923\\
0.125692067503251	0.093052158974376\\
0.122879513774241	0.0931457608654802\\
0.120912393202201	0.093238060256054\\
0.119422745738168	0.0933277040598274\\
0.118252155350029	0.0934141033604562\\
0.117296316251908	0.0934969835388925\\
0.116497691759121	0.0935771015865758\\
0.115799662926384	0.0936569207973493\\
0.115185198158437	0.0937359248145443\\
0.111214174417023	0.0945399374524934\\
0.109215168323888	0.0952994181438292\\
0.107822382950451	0.0960778329020644\\
0.106816530973209	0.0968873557192587\\
0.106083969980024	0.0977193158132013\\
0.105550300950939	0.0985794212016075\\
0.105172495087543	0.0994675469292449\\
0.104921959858263	0.100383626097969\\
0.104783061826401	0.101335154593886\\
0.104730900293759	0.102318706984575\\
};
\addlegendentry{WCRB ($K=1$)}

\addplot [color=mycolor2, densely dotted, line width=2.0pt]
  table[row sep=crcr]{%
0.137707668116726	0.0929118935698866\\
0.136846838049628	0.0929452144036222\\
0.136017843627891	0.0929804288188421\\
0.135218750990779	0.0930174654818698\\
0.134447581471313	0.0930562657664092\\
0.13370255154273	0.09309677568309\\
0.132982029591994	0.093138946021708\\
0.132284764774949	0.0931827195280599\\
0.131608850834164	0.0932280812056662\\
0.130953389266957	0.0932749759506937\\
0.130317204462981	0.0933233694379843\\
0.124817910709717	0.093883500951453\\
0.120899086937259	0.0944863921628044\\
0.117605946927391	0.0951784673178278\\
0.114781040023559	0.0959544139742411\\
0.112328558799844	0.0968112627895246\\
0.110191276968792	0.0977476873740165\\
0.108340734232879	0.0987636279346499\\
0.106776444888327	0.0998600808078692\\
0.105536048321296	0.101038971619326\\
0.104729089933781	0.102303233299158\\
};
\addlegendentry{WCM ($K=1$)}

\addplot [color=mycolor3, densely dotted, line width=2.0pt]
  table[row sep=crcr]{%
0.137707600336279	0.0929118899381638\\
0.137625382491875	0.0929142932402235\\
0.137526482016154	0.0929216583621338\\
0.137410295901031	0.0929342521961713\\
0.137276253360134	0.0929523419617807\\
0.137123876630561	0.0929761860362062\\
0.136952334305568	0.0930060802116067\\
0.136761649965585	0.0930422241473437\\
0.136551351230957	0.0930849134652983\\
0.136321397265766	0.093134396994818\\
0.136071244098507	0.0931907479661028\\
0.132466597160631	0.0941801275704443\\
0.127818555821408	0.0957252450837979\\
0.122401110628859	0.097717525598449\\
0.116966012116735	0.0997134359080259\\
0.112279881226114	0.101198270028581\\
0.108829515269526	0.101973322613998\\
0.106637260095203	0.102235391567727\\
0.105426430838976	0.102278371824349\\
0.104873232492036	0.102281484506814\\
0.104728926759788	0.102302992407712\\
};
\addlegendentry{WBF ($K=1$)}

\addplot [color=mycolor1, dashed, line width=2.0pt]
  table[row sep=crcr]{%
0.100771056988636	0.100445444560305\\
0.0974733611904515	0.100460667061682\\
0.095102608267812	0.100494656306167\\
0.0932976590075987	0.100537639980245\\
0.0918634752601958	0.100585266281622\\
0.0906881829813351	0.100635337920146\\
0.089705031656715	0.100686447807554\\
0.0888554020726498	0.100738638766791\\
0.0881075971486376	0.100791732012714\\
0.0874411037630687	0.100845518625433\\
0.086847298188996	0.100899326077806\\
0.0827146021955671	0.101490470344979\\
0.0801585488095344	0.102148589230116\\
0.0779607448723925	0.102995476366965\\
0.0759322017350362	0.104099312492658\\
0.0739930804166284	0.105555026762903\\
0.0720800262282712	0.107533055104814\\
0.0701498894124928	0.110349914057022\\
0.0681392531428259	0.114752951095956\\
0.0659962243789528	0.122992873598741\\
0.0641208974509642	0.151165658996798\\
};
\addlegendentry{WCRB ($K=2$)}

\addplot [color=mycolor2, dashed, line width=2.0pt]
  table[row sep=crcr]{%
0.100761410763011	0.100445389736233\\
0.099400785213614	0.100525258556117\\
0.0981212805420612	0.100614604345229\\
0.0969148986370815	0.100713261667122\\
0.0957741887356212	0.100820796471237\\
0.0946937064309009	0.100936828685986\\
0.0936681710300426	0.101061032077017\\
0.0926933148823108	0.101192779814347\\
0.0917644656498452	0.101332509901911\\
0.0908783133990363	0.101479686326012\\
0.0900316307043073	0.101634125025345\\
0.083218385212273	0.10355073205196\\
0.078855294777126	0.105790228765524\\
0.0754650574984189	0.108534914089046\\
0.07273909078477	0.11182420479348\\
0.0704983567444731	0.115731872227089\\
0.0686341369745584	0.120371470598895\\
0.0670809483702473	0.125910360850182\\
0.0658049598030488	0.13259536440541\\
0.0648032251262209	0.140799747725549\\
0.0641178038592577	0.151110318285503\\
};
\addlegendentry{WCM ($K=2$)}

\addplot [color=mycolor3, dashed, line width=2.0pt]
  table[row sep=crcr]{%
0.100761408511231	0.100445386785407\\
0.100462829781585	0.100453178474234\\
0.100153849409146	0.100476845788951\\
0.0998338950125332	0.100516805524326\\
0.0995013870009421	0.10057320065575\\
0.0991603838323559	0.100646497157976\\
0.0988097205797751	0.100736933141774\\
0.0984513092444917	0.100844840366354\\
0.0980857892111958	0.100970121680953\\
0.097715657294042	0.101113110215556\\
0.0973410270475782	0.101273853362737\\
0.0937702062789693	0.103776186423174\\
0.0911827447976084	0.106883942567912\\
0.0868213395656167	0.11061851573176\\
0.08007955138901	0.11592087098729\\
0.0739053262411722	0.123022291823995\\
0.0695331425452854	0.130818481208568\\
0.0667704194679617	0.137793244521436\\
0.0651604676700107	0.143167811926952\\
0.064351008627919	0.147305780902794\\
0.0641177342697953	0.151109758974223\\
};
\addlegendentry{WBF ($K=2$)}

\addplot [color=mycolor1, line width=2.0pt]
  table[row sep=crcr]{%
0.0827873031720473	0.106839978252764\\
0.0818947385382188	0.106841871111424\\
0.0810946444238936	0.106851017212691\\
0.0804055044780092	0.106864335053041\\
0.0797949868044433	0.106880862777541\\
0.0792462621964328	0.106900008260263\\
0.0787528510518854	0.106921188392837\\
0.078301897979017	0.106944173536233\\
0.0778876317776313	0.10696866809789\\
0.0775047115779013	0.106994476057877\\
0.0771487021778817	0.107021456670442\\
0.0744989378276438	0.107343488471397\\
0.0728382776934224	0.107709549410139\\
0.0714324009811981	0.108179279990219\\
0.0701310915395729	0.108801368056383\\
0.068857596743428	0.109652378417299\\
0.0675650693156507	0.110860217158332\\
0.0662144617155322	0.112669696506286\\
0.0647727293892009	0.115622544636856\\
0.0632136137929659	0.121363694629064\\
0.0618872332542553	0.140793205288059\\
};
\addlegendentry{WCRB ($K=3$)}

\addplot [color=mycolor2, line width=2.0pt]
  table[row sep=crcr]{%
0.0828147110606447	0.106838402426525\\
0.0822416424186669	0.106910495392562\\
0.0816887683918264	0.106986684774085\\
0.0811565378928959	0.107069791769705\\
0.0806420156123785	0.107156583433884\\
0.0801453213029607	0.107249090036872\\
0.0796650300301214	0.107346580112768\\
0.0792003038228014	0.107448980663237\\
0.0787502684273741	0.107556207548042\\
0.0783141393343669	0.107668179413936\\
0.0778911947973175	0.107784834820408\\
0.0742604658468092	0.109199294462832\\
0.0717190861331691	0.11082541456348\\
0.0696278373305535	0.112802837232242\\
0.067873863975644	0.115151531323355\\
0.0663858965874193	0.117912001542751\\
0.0651170683333172	0.12113874065297\\
0.0640380922875498	0.124906650092137\\
0.0631354908888918	0.129327367929788\\
0.062410795956803	0.134551175830391\\
0.0618870934369973	0.140782274285841\\
};
\addlegendentry{WCM ($K=3$)}

\addplot [color=mycolor3, line width=2.0pt]
  table[row sep=crcr]{%
0.0828148269500874	0.106838394863626\\
0.0827349553142833	0.106846507981573\\
0.0826543678214041	0.106871271545822\\
0.0825714359008368	0.106913349416308\\
0.0824869865728083	0.106973432445834\\
0.0824006148826319	0.107052235516644\\
0.0823123310356481	0.107150505620974\\
0.0822219321134644	0.10726901590645\\
0.0821299188762316	0.107408581881558\\
0.082035436389029	0.107570027983045\\
0.0819392905998717	0.107754243181616\\
0.0809012674027534	0.11104888307198\\
0.0800409101012114	0.116313123908518\\
0.0785548186923971	0.121467953440025\\
0.0746895661266249	0.125081954621801\\
0.0703558841242184	0.129347602733919\\
0.0668938739455238	0.133612944027667\\
0.0644646842230661	0.13660054071303\\
0.0629321214697749	0.138263991180524\\
0.0621245077274641	0.139402816959419\\
0.0618869762754298	0.140782042671297\\
};
\addlegendentry{WBF ($K=3$)}

\end{axis}
\end{tikzpicture}%
  \vspace{-1cm}
  \centerline{(a) Digital \ac{fdb}} \medskip
\end{minipage}
\hfill
\begin{minipage}[b]{0.48\linewidth}
  \centering
\definecolor{mycolor1}{rgb}{0.49400,0.18400,0.55600}%
\definecolor{mycolor2}{rgb}{0.30100,0.74500,0.93300}%
\definecolor{mycolor3}{rgb}{0.46600,0.67400,0.18800}%
\begin{tikzpicture}

\begin{axis}[%
width=72mm,
height=40mm,
at={(0mm, 0mm)},
scale only axis,
xmin=0.1,
xmax=0.4,
xticklabel style = {font=\color{white!15!black},font=\footnotesize},
xlabel style={font=\footnotesize, xshift=2mm},
xlabel={$\sqrt{\text{CRB}}$ of \ac{bp} [m]},
ymin=0.1,
ymax=0.18,
yminorticks=true,
yticklabel style = {font=\color{white!15!black},font=\footnotesize},
ylabel style={font=\footnotesize, yshift=0mm,font=\footnotesize},
ylabel={$\sqrt{\text{CRB}}$ of \ac{ms} [m]},
axis background/.style={fill=white},
xmajorgrids,
ymajorgrids,
yminorgrids,
legend style={font=\scriptsize, at={(1.0,1.34)}, anchor=north east, legend cell align=left, align=left, fill = white, fill opacity=0.9, legend columns = 3}
]
\addplot [color=mycolor1, densely dotted, line width=2.0pt]
  table[row sep=crcr]{%
0.395104193827843	0.100077202985003\\
0.22384438924392	0.100553833031306\\
0.213092391916679	0.100892979942845\\
0.20705359399624	0.101209906162207\\
0.202824217306895	0.10151770700152\\
0.199590890301966	0.101817858362107\\
0.196966273695873	0.10211391583841\\
0.194752032035026	0.102408197962641\\
0.192883697856977	0.102694090993391\\
0.19123233485809	0.102980118517459\\
0.18977367321322	0.103262600171876\\
0.180577978981884	0.106003080805183\\
0.176110604780907	0.108379529828788\\
0.173201132551472	0.110681521526318\\
0.171071362692515	0.113065080964428\\
0.169591276685145	0.115369201174042\\
0.168507419867577	0.117721513135822\\
0.167761412055053	0.120047731667489\\
0.167258552025783	0.122463961698574\\
0.166978556873084	0.12491629244756\\
0.1668906886939	0.127398883999782\\
};
\addlegendentry{WCRB ($K=1$)}

\addplot [color=mycolor2, densely dotted, line width=2.0pt]
  table[row sep=crcr]{%
0.395389959832022	0.100077193730725\\
0.374260878375729	0.100081536303319\\
0.356738428537646	0.100094299445315\\
0.341910121792013	0.100115108945953\\
0.329156667415023	0.100143621654536\\
0.318041902846777	0.100179518717809\\
0.308245627914002	0.100222516307789\\
0.299530773864323	0.100272347662124\\
0.291712629391636	0.100328764385098\\
0.284651015735682	0.100391538897019\\
0.278232381818676	0.100460465182323\\
0.23554832282461	0.101452380567572\\
0.214245309730057	0.102735755284569\\
0.200016247189018	0.104355345693698\\
0.189842070627833	0.106302174164379\\
0.182296095364113	0.108589683763497\\
0.176622933288447	0.111252304516945\\
0.172401797114001	0.114348434666093\\
0.169406835579685	0.117968352456012\\
0.167551948215015	0.122249690270961\\
0.166890687779902	0.12740686435925\\
};
\addlegendentry{WCM ($K=1$)}

\addplot [color=mycolor3, densely dotted, line width=2.0pt]
  table[row sep=crcr]{%
0.395419121110602	0.100077193744087\\
0.386155488228746	0.100079272829549\\
0.377254450313431	0.100085669384028\\
0.368783191498537	0.100096320776045\\
0.360682824670438	0.100111355276757\\
0.352896719925942	0.100130978851631\\
0.345389343789738	0.100155513670539\\
0.338242761029811	0.10018467183696\\
0.331416904437003	0.100218435551265\\
0.324855104752092	0.100257244656494\\
0.318590760029876	0.100300793486794\\
0.268160602548939	0.101024949003476\\
0.237304824981819	0.102135745025532\\
0.21510564030936	0.103711216051484\\
0.198984160055374	0.105762824399914\\
0.187286649210537	0.108291445337092\\
0.178935994977128	0.111287702010799\\
0.173183315042022	0.11473308308676\\
0.169502251280143	0.118597979690254\\
0.167503206254898	0.122841976731761\\
0.166890687785002	0.127406779482761\\
};
\addlegendentry{WBF ($K=1$)}

\addplot [color=mycolor1, dashed, line width=2.0pt]
  table[row sep=crcr]{%
0.270206599669279	0.111419104298645\\
0.175699746247648	0.111683452742173\\
0.16706517737561	0.111875192491972\\
0.162407452793231	0.112047911858389\\
0.159227111726514	0.112212050651189\\
0.156775858413034	0.112373911470014\\
0.15479108418648	0.112533513363678\\
0.153088753366369	0.112694908995115\\
0.151599209032321	0.112857778116914\\
0.150278215168038	0.113021514465226\\
0.149064930736696	0.113189839865515\\
0.140672718161723	0.115006110741971\\
0.135668144255905	0.116920951707981\\
0.131805580819229	0.119105185976892\\
0.12857016096169	0.121659193598493\\
0.125799472213186	0.124652123071163\\
0.123376993919236	0.128240854627074\\
0.121227585908952	0.132726167353124\\
0.119306471316193	0.138753718536419\\
0.117571407071508	0.148392709912595\\
0.116380255853106	0.172833437610405\\
};
\addlegendentry{WCRB ($K=2$)}

\addplot [color=mycolor2, dashed, line width=2.0pt]
  table[row sep=crcr]{%
0.271029918169218	0.111419099541718\\
0.257802645283496	0.111426389548394\\
0.246676029741197	0.111447739214209\\
0.237174876426387	0.111482440258554\\
0.228944121680489	0.111529888278548\\
0.221711414897428	0.111589529411995\\
0.215305205523416	0.111660852875892\\
0.209573286738691	0.111743431845644\\
0.20440532608384	0.111836873871881\\
0.199717312519254	0.111940817242007\\
0.195440618805137	0.112054939537806\\
0.166566932526241	0.113706077146588\\
0.151812620691225	0.115882027268024\\
0.141769737690422	0.118702804018523\\
0.134448354774444	0.122211395227527\\
0.128896451531715	0.12650914757878\\
0.12460415758792	0.131771160082763\\
0.121285158483911	0.138285825182137\\
0.118789890172846	0.146542593687117\\
0.117089238312407	0.157440779851641\\
0.116380263091616	0.172872067411436\\
};
\addlegendentry{WCM ($K=2$)}

\addplot [color=mycolor3, dashed, line width=2.0pt]
  table[row sep=crcr]{%
0.271028312511391	0.111419099548246\\
0.267259540559567	0.111422321115082\\
0.263426145103896	0.111432126030733\\
0.259528211187331	0.111448862933802\\
0.255612034209467	0.111472586093858\\
0.251666014099122	0.111503757866415\\
0.247712787283004	0.111542681004413\\
0.243754855512317	0.111589761998957\\
0.239847837575678	0.111644450112302\\
0.235954294834062	0.1117080586216\\
0.232110401664474	0.111780141816732\\
0.197345049207218	0.113031251437617\\
0.173353809905408	0.115069779343211\\
0.155381702924326	0.118113866206292\\
0.142196806798179	0.122267089118916\\
0.13266689690964	0.127615056414302\\
0.125927849378313	0.134218179959372\\
0.121338341623166	0.142107456256236\\
0.118429951048965	0.151269464120515\\
0.11685991759493	0.161602444088901\\
0.116380264199625	0.172877365230408\\
};
\addlegendentry{WBF ($K=2$)}

\addplot [color=mycolor1, line width=2.0pt]
  table[row sep=crcr]{%
0.151409576670851	0.12136040303444\\
0.136090255802373	0.121420509575487\\
0.131657083063191	0.121492028936449\\
0.129063023812588	0.12156261083609\\
0.127365712689238	0.12162707767165\\
0.126058671229292	0.121690953647524\\
0.125009232978169	0.121753926753975\\
0.124180270510127	0.121812784661968\\
0.123463470378846	0.121871753916279\\
0.122842975116097	0.121929912242517\\
0.122283151205337	0.121989290534354\\
0.118841831944344	0.122564976082608\\
0.116979973033949	0.123136946498038\\
0.115590103343297	0.123788075209681\\
0.114446223744868	0.12456094828032\\
0.113446498294809	0.125515177753491\\
0.112518449557615	0.12676856030202\\
0.111596367758191	0.128579842268232\\
0.110648420108789	0.131473039781555\\
0.10969996996944	0.13674794885341\\
0.109069318365017	0.149740199136381\\
};
\addlegendentry{WCRB ($K=3$)}

\addplot [color=mycolor2, line width=2.0pt]
  table[row sep=crcr]{%
0.151193379212139	0.12136013305539\\
0.149394921084051	0.121364078385344\\
0.147703151080918	0.121374883279153\\
0.146109277978573	0.121392340405663\\
0.144604578385202	0.121416256510264\\
0.143182178062251	0.121446461396254\\
0.141834644108268	0.121482792799344\\
0.140555471600291	0.121525113806047\\
0.13934034094611	0.121573263149391\\
0.138184044594614	0.121627122605596\\
0.137082279328581	0.121686573244467\\
0.128363836103164	0.122568446584647\\
0.123001228939265	0.123757191899506\\
0.119031055603372	0.125307594837705\\
0.116023206914595	0.127224470524315\\
0.113725100012195	0.129532379783583\\
0.111980304353551	0.132276775057683\\
0.110689659757533	0.135528997520174\\
0.109791742931161	0.139396520969104\\
0.109253697091204	0.144042022312564\\
0.10906933393495	0.149720019222938\\
};
\addlegendentry{WCM ($K=3$)}

\addplot [color=mycolor3, line width=2.0pt]
  table[row sep=crcr]{%
0.15119573556747	0.121360133067936\\
0.150273386083749	0.121361999360182\\
0.149353456020385	0.121367710536426\\
0.148446302518538	0.121377247221163\\
0.147539952378807	0.121390859799724\\
0.14664427164982	0.121408503776366\\
0.145754914983855	0.121430357555373\\
0.1448739866736	0.121456448646755\\
0.143998353089831	0.12148689755998\\
0.14314140148093	0.121521455138772\\
0.142282042519105	0.121561018631031\\
0.134324104453157	0.122219423031265\\
0.12823480172086	0.123248911039208\\
0.123116506554864	0.12473965486534\\
0.118924906322121	0.126726938732627\\
0.115595965431284	0.129234905786808\\
0.113049641205445	0.132277003907385\\
0.111201769724592	0.135861459736386\\
0.109972229381973	0.139978083011649\\
0.109284781294361	0.144608978916523\\
0.109069334025544	0.149718914807069\\
};
\addlegendentry{WBF ($K=3$)}

\end{axis}
\end{tikzpicture}%
  \vspace{-1cm}
  \centerline{(b) Digital \ac{cpa}} \medskip
\end{minipage}

% \vspace{0.3cm} % Adjust vertical spacing between rows if needed

\begin{minipage}[b]{0.48\linewidth}
  \centering
\definecolor{mycolor1}{rgb}{1.00000,0.00000,1.00000}%
\begin{tikzpicture}

\begin{axis}[%
width=72mm,
height=40mm,
at={(0mm, 0mm)},
scale only axis,
xmin=0.07,
xmax=0.17,
xticklabel style = {font=\color{white!15!black},font=\footnotesize},
xlabel style={font=\footnotesize, xshift=2mm},
xlabel={$\sqrt{\text{CRB}}$ of \ac{bp} [m]},
ymin=0.1,
ymax=0.19,
yminorticks=true,
yticklabel style = {font=\color{white!15!black},font=\footnotesize},
ylabel style={font=\footnotesize, yshift=0mm,font=\footnotesize},
ylabel={$\sqrt{\text{CRB}}$ of \ac{ms} [m]},
axis background/.style={fill=white},
xmajorgrids,
ymajorgrids,
yminorgrids,
legend style={font=\scriptsize, legend cell align=left, align=left, draw=white!15!black}
]
\addplot [color=mycolor1, line width=2.0pt, mark size=5.0pt, mark=x, mark options={solid, mycolor1}]
  table[row sep=crcr]{%
0.160548628947769	0.110178404603328\\
0.156365143158451	0.11010192085925\\
0.155013183780561	0.110202165192057\\
0.152621586805929	0.110376518699804\\
0.142210370694563	0.110398528901851\\
0.134628758113928	0.111405889125425\\
0.128298576213334	0.112118251189113\\
0.124160582570882	0.113283253404407\\
0.125126422398137	0.115817446319473\\
};
\addlegendentry{$K=1$}

\addplot [color=mycolor1, line width=2.0pt, mark size=3.5pt, mark=square, mark options={solid, mycolor1}]
  table[row sep=crcr]{%
0.112795898545332	0.120757970952782\\
0.111145008865567	0.120425701843349\\
0.109003071534062	0.12022743959123\\
0.10609448723083	0.120910577628717\\
0.0931245466143222	0.124549597712761\\
0.085475473591266	0.130539667168443\\
0.0791794265390914	0.142479467475325\\
0.0770489517951459	0.162548236997228\\
0.0755120439312642	0.186424711938014\\
};
\addlegendentry{$K=2$}

\addplot [color=mycolor1, line width=2.0pt,  mark size=3.3pt, mark=triangle, mark options={solid, rotate=180, mycolor1}]
  table[row sep=crcr]{%
0.105406791282052	0.133447299831971\\
0.105607035074839	0.132734386107236\\
0.102403275866833	0.133639694433727\\
0.0994891662570688	0.13282970450886\\
0.0909988717010226	0.135549750636676\\
0.0843693845324138	0.140405466235811\\
0.0794681343642414	0.150434773319429\\
0.0771768879297024	0.163049614395785\\
0.0770175107959225	0.183990802658981\\
};
\addlegendentry{$K=3$}

\end{axis}
\end{tikzpicture}%
  \vspace{-1cm}
  \centerline{(c) Analog \ac{fdb}} \medskip
\end{minipage}
\hfill
\begin{minipage}[b]{0.48\linewidth}
  \centering
\definecolor{mycolor1}{rgb}{0.85000,0.32500,0.09800}%
\begin{tikzpicture}

\begin{axis}[%
width=72mm,
height=40mm,
at={(0mm, 0mm)},
scale only axis,
xmin=0,
xmax=1.2,
xticklabel style = {font=\color{white!15!black},font=\footnotesize},
xlabel style={font=\footnotesize, xshift=2mm},
xlabel={$\sqrt{\text{CRB}}$ of \ac{bp} [m]},
ymin=0.1,
ymax=0.17,
yminorticks=true,
yticklabel style = {font=\color{white!15!black},font=\footnotesize},
ylabel style={font=\footnotesize, yshift=0mm,font=\footnotesize},
ylabel={$\sqrt{\text{CRB}}$ of \ac{ms} [m]},
axis background/.style={fill=white},
xmajorgrids,
ymajorgrids,
yminorgrids,
legend style={font=\scriptsize, legend cell align=left, align=left, draw=white!15!black}
]
\addplot [color=blue, line width=2.0pt, mark size=5.0pt, mark=x, mark options={solid, blue}]
  table[row sep=crcr]{%
1.10368479937271	0.100254632690453\\
1.03473873695579	0.100335286109379\\
0.979792681649899	0.100473648315009\\
0.933727171292366	0.100664702870688\\
0.744432416436734	0.103012335240438\\
0.650258200570787	0.106407489647209\\
0.586179349186534	0.110920556590237\\
0.539078793376118	0.116652933189752\\
0.502844166592084	0.123875703640089\\
};
\addlegendentry{$K=1$}

\addplot [color=blue, line width=2.0pt, mark size=3.5pt, mark=square, mark options={solid, blue}]
  table[row sep=crcr]{%
0.330051183926678	0.111686616268348\\
0.3183567820343	
0.111837041864306\\
0.30794917828808	0.112082608802779\\
0.298627318415417	0.112415078767253\\
0.254433954359274	0.116507112984546\\
0.229556577163391	0.122775319972992\\
0.212116720095021	0.131929821481024\\
0.199748211985531	0.145212317664692\\
0.191626843675505	0.165491176749529\\
};
\addlegendentry{$K=2$}

\addplot [color=blue, line width=2.0pt, mark size=3.3pt, mark=triangle, mark options={solid, rotate=180, blue}]
  table[row sep=crcr]{%
0.189345264896284	0.121595027139269\\
0.18631975326895	0.121670465368634\\
0.183523800110896	0.121795164390795\\
0.180934139365864	0.121966802904604\\
0.167775950236694	0.124185224252799\\
0.160147061729113	0.127667052024426\\
0.155131330293742	0.132646322687839\\
0.152103060533926	0.139428025987481\\
0.150801179720571	0.148609033232143\\
};
\addlegendentry{$K=3$}

\end{axis}
\end{tikzpicture}%
  \vspace{-1cm}
  \centerline{(d) Analog \ac{cpa}} \medskip
\end{minipage}

\vspace{-0.4cm}
\caption{
Comparison of the tradeoff in terms of the square root of the \ac{crb} between \ac{bp} and \ac{ms} across various target numbers: (a) Digital \ac{fdb}; (b) Digital \ac{cpa}; (c) Analog \ac{fdb}; and (d) Analog \ac{cpa}.}
\label{tradeoff-diff-scheme}
\end{figure*}

\subsection{Results and Discussion}
\subsubsection{Convergence}
The proposed digital beamforming approaches (as well as Analog-\ac{cpa} approach) are obtained either by \ac{sdp} or \ac{qcqp} and are solved via CVX with self-contained convergence. The proposed Analog-\ac{fdb} approach is designed under the \ac{ao} framework. Specifically, the convergence of objective value of $f(\boldsymbol{\rho},\boldsymbol{\Phi})$ is shown in Fig. \ref{converence-plot}. We observe that the objective value saturates within six iterations, validating the convergence of solving Analog-\ac{fdb} via Algorithm \ref{sqp-anlog-bf}. 
%The convergence of Analog-\ac{fdb} is determined by the analog codebook construction phase, i.e., the \ac{ao} process summarized in Algorithm \ref{analog-codeword-ao}. The objective values of constructing the four analog codewords (calculated by \eqref{manalog-code-construct-obj}) over iterations are shown in Fig. \ref{converence-plot}(b), which also saturate within six iterations, validating the convergence of Algorithm \ref{analog-codeword-ao} as well as the proposed Analog-\ac{cpa} approach.

\subsubsection{Beampatterns}
The core rationale behind the adopted Analog-\ac{cpa} approach is to replicate the digital codebook through an analog equivalent. To illustrate this, Fig. \ref{analog-codebook-compare} compares the beampatterns of various digital-analog pairs. As shown, the dashed lines represent the beampatterns of the generated analog codewords and closely match the solid lines, which represent those of the original digital codewords. This strong alignment confirms that the analog codebook construction method\cite{tranter2017tsp}, effectively preserves the spatial signatures, thereby substantiating the effectiveness of the Analog-\ac{cpa} scheme, which we further elaborate on subsequently.

In Fig. \ref{bp}, we present the 2D beampatterns for various approaches across different scenarios, illustrating the spatial behavior and characteristics of the proposed beamforming techniques under extreme conditions. This comparison provides insights into the performance tradeoff between \ac{bp} and \ac{ms}, which we will further analyze in detail later. The top row shows results for \ac{bp} ($\alpha = 1$), while the bottom row corresponds to \ac{ms} ($\alpha = 0$). Within each row, the outcomes for the Digital-\ac{fdb}, Analog-\ac{fdb}, Digital-\ac{cpa}, and Analog-\ac{cpa} schemes are shown sequentially from left to right.

In general, for \ac{bp}, all schemes direct strong beams toward the \ac{ue}, with variations in power allocated to beams aimed at the targets. This occurs because the \ac{bs} serves as the primary anchor in \ac{bp}, while the targets act as secondary anchors, with their positions being estimated concurrently. The information provided about the \ac{ue}'s position varies based on relative locations, which leads to a power adjustment across beams illuminating the targets to maximize \ac{ue} positioning accuracy. In \ac{ms}, accurate positioning of both the targets and the \ac{ue} is crucial for optimal performance. Due to the \ac{ue}'s relatively small \ac{rcs}, stronger beams are consistently directed toward it in all schemes to maintain balanced positioning accuracy across all targets.

Additionally, beams generated by \ac{fdb}-based approaches are broader compared to those from \ac{cpa}-based approaches. This is due to a larger design \acp{dof} in \ac{fdb}-based methods, which enable the synthesis of beams directed around the targets from a predetermined codebook, thus enhancing spatial information extraction for superior \ac{bp} and/or \ac{ms} performance. Conversely, \ac{cpa}-based approaches, constrained by limited design \acp{dof} due to their power allocation nature, can only transmit beams along predetermined directions. Furthermore, both the digital approaches and their analog counterparts show alignment in their beampatterns, demonstrating the potential efficacy of the proposed analog beamforming approaches.

% \begin{figure}[htb]
% % \vspace{-0.1cm}
% \centering
% \begin{minipage}[b]{0.98\linewidth}
% \vspace{-0.3cm}
%   \centering
%     \include{Figures/sp2}
%     \vspace{-1.cm}
%   \centerline{(a) $K=2$} \medskip
% \end{minipage}
% \hfill
% \begin{minipage}[b]{0.98\linewidth}
% \vspace{-0.3cm}
%   \centering
%     \include{Figures/sp3}
%     \vspace{-1.cm}
%   \centerline{(b) $K=3$} \medskip
% \end{minipage}
% \vspace{-0.4cm}
% \caption{
% Tradeoff (in terms of square root of \ac{crb}) between \ac{bp} and \ac{ms}: (a) $K=2$ (targets 1 and 2); (b) $K=3$ (targets 1, 2, and 3).}
% \label{tradeoff}
% % \vspace{-0.2cm}
% \end{figure}

% \begin{figure}[t]
% \centering 
% \centerline{\input{Figures/sp3}}
% \caption{Comparison of the tradeoff in terms of the square root of \ac{crb} between \ac{bp} and \ac{ms} across various schemes.}
% % \vspace{-5mm}
% \label{tradeoff}
% \end{figure}

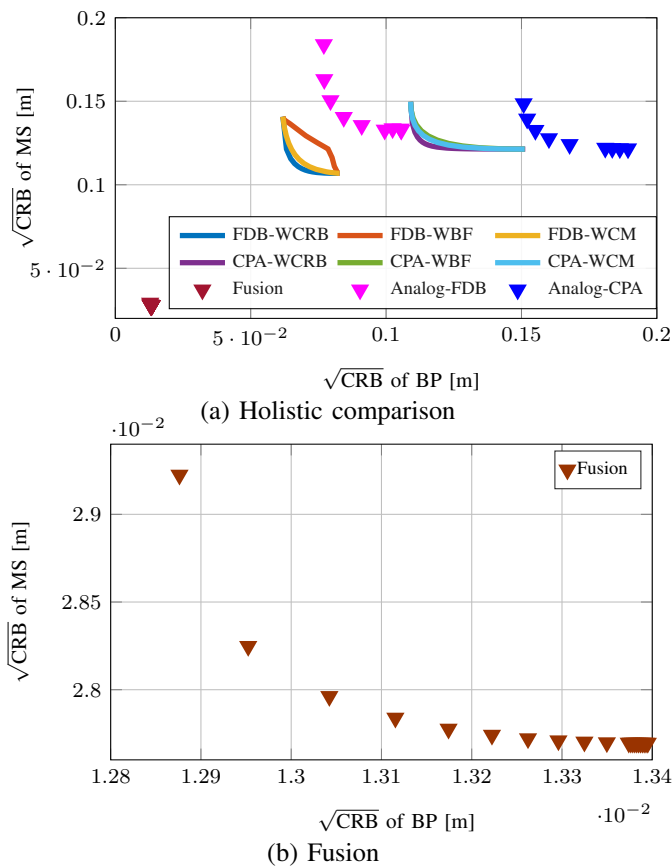
\begin{figure}[t]
% \vspace{-0.1cm}
\centering
\begin{minipage}[b]{0.98\linewidth}
\vspace{-0.3cm}
  \centering
\definecolor{mycolor1}{rgb}{0.00000,0.44700,0.74100}%
\definecolor{mycolor2}{rgb}{0.85000,0.32500,0.09800}%
\definecolor{mycolor3}{rgb}{0.92900,0.69400,0.12500}%
\definecolor{mycolor4}{rgb}{0.49400,0.18400,0.55600}%
\definecolor{mycolor5}{rgb}{0.46667,0.67451,0.18824}%
\definecolor{mycolor6}{rgb}{0.30100,0.74500,0.93300}%
\definecolor{mycolor7}{rgb}{0.63500,0.07800,0.18400}%
\definecolor{mycolor8}{rgb}{1.00000,0.00000,1.00000}%
\begin{tikzpicture}
% [font=\footnotesize, spy using outlines={rectangle, magnification=10, size=1cm, connect spies}]
\begin{axis}[%
width=72mm,
height=40mm,
at={(0mm, 0mm)},
scale only axis,
xmin=0,
xmax=0.2,
xticklabel style = {font=\color{white!15!black},font=\footnotesize},
xlabel style={font=\footnotesize, xshift=2mm},
xlabel={$\sqrt{\text{CRB}}$ of \ac{bp} [m]},
ymin=0.02,
ymax=0.2,
yminorticks=true,
yticklabel style = {font=\color{white!15!black},font=\footnotesize},
ylabel style={font=\footnotesize, yshift=0mm,font=\footnotesize},
ylabel={$\sqrt{\text{CRB}}$ of \ac{ms} [m]},
axis background/.style={fill=white},
xmajorgrids,
ymajorgrids,
yminorgrids,
legend style={font=\scriptsize, at={(1,0.34)}, anchor=north east, legend cell align=left, align=left, fill = white, fill opacity=0.9, legend columns = 3}
]
\addplot [color=mycolor1, line width=2.0pt]
  table[row sep=crcr]{%
0.0827873031720473	0.106839978252764\\
0.0818947385382188	0.106841871111424\\
0.0810946444238936	0.106851017212691\\
0.0804055044780092	0.106864335053041\\
0.0797949868044433	0.106880862777541\\
0.0792462621964328	0.106900008260263\\
0.0787528510518854	0.106921188392837\\
0.078301897979017	0.106944173536233\\
0.0778876317776313	0.10696866809789\\
0.0775047115779013	0.106994476057877\\
0.0771487021778817	0.107021456670442\\
0.0744989378276438	0.107343488471397\\
0.0728382776934224	0.107709549410139\\
0.0714324009811981	0.108179279990219\\
0.0701310915395729	0.108801368056383\\
0.068857596743428	0.109652378417299\\
0.0675650693156507	0.110860217158332\\
0.0662144617155322	0.112669696506286\\
0.0647727293892009	0.115622544636856\\
0.0632136137929659	0.121363694629064\\
0.0618872332542553	0.140793205288059\\
};
\addlegendentry{FDB-WCRB}

\addplot [color=mycolor2, line width=2.0pt]
  table[row sep=crcr]{%
0.0828148269500874	0.106838394863626\\
0.0827349553142833	0.106846507981573\\
0.0826543678214041	0.106871271545822\\
0.0825714359008368	0.106913349416308\\
0.0824869865728083	0.106973432445834\\
0.0824006148826319	0.107052235516644\\
0.0823123310356481	0.107150505620974\\
0.0822219321134644	0.10726901590645\\
0.0821299188762316	0.107408581881558\\
0.082035436389029	0.107570027983045\\
0.0819392905998717	0.107754243181616\\
0.0809012674027534	0.11104888307198\\
0.0800409101012114	0.116313123908518\\
0.0785548186923971	0.121467953440025\\
0.0746895661266249	0.125081954621801\\
0.0703558841242184	0.129347602733919\\
0.0668938739455238	0.133612944027667\\
0.0644646842230661	0.13660054071303\\
0.0629321214697749	0.138263991180524\\
0.0621245077274641	0.139402816959419\\
0.0618869762754298	0.140782042671297\\
};
\addlegendentry{FDB-WBF}

\addplot [color=mycolor3, line width=2.0pt]
  table[row sep=crcr]{%
0.0828147110606447	0.106838402426525\\
0.0822416424186669	0.106910495392562\\
0.0816887683918264	0.106986684774085\\
0.0811565378928959	0.107069791769705\\
0.0806420156123785	0.107156583433884\\
0.0801453213029607	0.107249090036872\\
0.0796650300301214	0.107346580112768\\
0.0792003038228014	0.107448980663237\\
0.0787502684273741	0.107556207548042\\
0.0783141393343669	0.107668179413936\\
0.0778911947973175	0.107784834820408\\
0.0742604658468092	0.109199294462832\\
0.0717190861331691	0.11082541456348\\
0.0696278373305535	0.112802837232242\\
0.067873863975644	0.115151531323355\\
0.0663858965874193	0.117912001542751\\
0.0651170683333172	0.12113874065297\\
0.0640380922875498	0.124906650092137\\
0.0631354908888918	0.129327367929788\\
0.062410795956803	0.134551175830391\\
0.0618870934369973	0.140782274285841\\
};
\addlegendentry{FDB-WCM}

\addplot [color=mycolor4, line width=2.0pt]
  table[row sep=crcr]{%
0.151409576670851	0.12136040303444\\
0.136090255802373	0.121420509575487\\
0.131657083063191	0.121492028936449\\
0.129063023812588	0.12156261083609\\
0.127365712689238	0.12162707767165\\
0.126058671229292	0.121690953647524\\
0.125009232978169	0.121753926753975\\
0.124180270510127	0.121812784661968\\
0.123463470378846	0.121871753916279\\
0.122842975116097	0.121929912242517\\
0.122283151205337	0.121989290534354\\
0.118841831944344	0.122564976082608\\
0.116979973033949	0.123136946498038\\
0.115590103343297	0.123788075209681\\
0.114446223744868	0.12456094828032\\
0.113446498294809	0.125515177753491\\
0.112518449557615	0.12676856030202\\
0.111596367758191	0.128579842268232\\
0.110648420108789	0.131473039781555\\
0.10969996996944	0.13674794885341\\
0.109069318365017	0.149740199136381\\
};
\addlegendentry{CPA-WCRB}

\addplot [color=mycolor5, line width=2.0pt]
  table[row sep=crcr]{%
0.15119573556747	0.121360133067936\\
0.150273386083749	0.121361999360182\\
0.149353456020385	0.121367710536426\\
0.148446302518538	0.121377247221163\\
0.147539952378807	0.121390859799724\\
0.14664427164982	0.121408503776366\\
0.145754914983855	0.121430357555373\\
0.1448739866736	0.121456448646755\\
0.143998353089831	0.12148689755998\\
0.14314140148093	0.121521455138772\\
0.142282042519105	0.121561018631031\\
0.134324104453157	0.122219423031265\\
0.12823480172086	0.123248911039208\\
0.123116506554864	0.12473965486534\\
0.118924906322121	0.126726938732627\\
0.115595965431284	0.129234905786808\\
0.113049641205445	0.132277003907385\\
0.111201769724592	0.135861459736386\\
0.109972229381973	0.139978083011649\\
0.109284781294361	0.144608978916523\\
0.109069334025544	0.149718914807069\\
};
\addlegendentry{CPA-WBF}

\addplot [color=mycolor6, line width=2.0pt]
  table[row sep=crcr]{%
0.151193379212139	0.12136013305539\\
0.149394921084051	0.121364078385344\\
0.147703151080918	0.121374883279153\\
0.146109277978573	0.121392340405663\\
0.144604578385202	0.121416256510264\\
0.143182178062251	0.121446461396254\\
0.141834644108268	0.121482792799344\\
0.140555471600291	0.121525113806047\\
0.13934034094611	0.121573263149391\\
0.138184044594614	0.121627122605596\\
0.137082279328581	0.121686573244467\\
0.128363836103164	0.122568446584647\\
0.123001228939265	0.123757191899506\\
0.119031055603372	0.125307594837705\\
0.116023206914595	0.127224470524315\\
0.113725100012195	0.129532379783583\\
0.111980304353551	0.132276775057683\\
0.110689659757533	0.135528997520174\\
0.109791742931161	0.139396520969104\\
0.109253697091204	0.144042022312564\\
0.10906933393495	0.149720019222938\\
};
\addlegendentry{CPA-WCM}

\addplot [color=mycolor7, line width=2.0pt, only marks, mark=triangle, mark options={solid, rotate=180, mycolor7}]
  table[row sep=crcr]{%
0.0133950304039107	0.0276957154820386\\
0.01339305020096	0.0276957197167439\\
0.0133910439646264	0.0276957343233446\\
0.0133890108909637	0.0276957599155256\\
0.0133869504635885	0.027695795759008\\
0.0133848616873925	0.0276958431536227\\
0.0133827444544839	0.0276959026949236\\
0.0133805985157348	0.0276959806300394\\
0.0133784213197911	0.0276960644610205\\
0.0133762140272694	0.0276961619121859\\
0.0133739759929848	0.0276962746826357\\
0.0133497197436697	0.0276983734351541\\
0.0133248315140289	0.027702290380577\\
0.0132960439370536	0.0277092706815651\\
0.0132623574367439	0.0277211786601774\\
0.0132224119230675	0.0277413341315904\\
0.0131743170346945	0.0277760745183204\\
0.0131154312153793	0.0278387739725025\\
0.0130422721667354	0.0279620545232888\\
0.0129523819146622	0.0282476934940976\\
0.0128762310555204	0.0292243410158573\\
};
\addlegendentry{Fusion}

\addplot [color=mycolor8, line width=2.0pt, only marks, mark=triangle, mark options={solid, rotate=180, mycolor8}]
  table[row sep=crcr]{%
0.105406791282052	0.133447299831971\\
0.105607035074839	0.132734386107236\\
0.102403275866833	0.133639694433727\\
0.0994891662570688	0.13282970450886\\
0.0909988717010226	0.135549750636676\\
0.0843693845324138	0.140405466235811\\
0.0794681343642414	0.150434773319429\\
0.0771768879297024	0.163049614395785\\
0.0770175107959225	0.183990802658981\\
};
\addlegendentry{Analog-FDB}

\addplot [color=blue, line width=2.0pt, only marks, mark=triangle, mark options={solid, rotate=180, blue}]
  table[row sep=crcr]{%
0.189345264896284	0.121595027139269\\
0.18631975326895	0.121670465368634\\
0.183523800110896	0.121795164390795\\
0.180934139365864	0.121966802904604\\
0.167775950236694	0.124185224252799\\
0.160147061729113	0.127667052024426\\
0.155131330293742	0.132646322687839\\
0.152103060533926	0.139428025987481\\
0.150801179720571	0.148609033232143\\
};
\addlegendentry{Analog-CPA}

% \begin{scope}
%     \spy[black, size=2.5cm] on (0.5, 0.3) in node [fill=none] at (4.5, 1);
% \end{scope}

\end{axis}
\end{tikzpicture}%
    \vspace{-1.cm}
  \centerline{(a) Holistic comparison} \medskip
\end{minipage}
\hfill
\begin{minipage}[b]{0.98\linewidth}
\vspace{-0.3cm}
  \centering
\definecolor{mycolor1}{rgb}{0.60000,0.20000,0.00000}%
\begin{tikzpicture}

\begin{axis}[%
width=72mm,
height=42mm,
at={(0in,0in)},
scale only axis,
xmin=0.0128,
xmax=0.0134,
xticklabel style = {font=\color{white!15!black},font=\footnotesize},
xlabel style={font=\footnotesize, xshift=2mm},
xlabel={$\sqrt{\text{CRB}}$ of \ac{bp} [m]},
ymin=0.0276,
ymax=0.0294,
yticklabel style = {font=\color{white!15!black},font=\footnotesize},
ylabel style={font=\footnotesize, yshift=0mm},
ylabel={$\sqrt{\text{CRB}}$ of \ac{ms} [m]},
axis background/.style={fill=white},
xmajorgrids,
ymajorgrids,
legend style={font=\scriptsize, legend cell align=left, align=left, draw=white!15!black}
]
\addplot [color=mycolor1, line width=2.0pt, only marks, mark=triangle, mark options={solid, rotate=180, mycolor1}]
  table[row sep=crcr]{%
0.0133950304039107	0.0276957154820386\\
0.01339305020096	0.0276957197167439\\
0.0133910439646264	0.0276957343233446\\
0.0133890108909637	0.0276957599155256\\
0.0133869504635885	0.027695795759008\\
0.0133848616873925	0.0276958431536227\\
0.0133827444544839	0.0276959026949236\\
0.0133805985157348	0.0276959806300394\\
0.0133784213197911	0.0276960644610205\\
0.0133762140272694	0.0276961619121859\\
0.0133739759929848	0.0276962746826357\\
0.0133497197436697	0.0276983734351541\\
0.0133248315140289	0.027702290380577\\
0.0132960439370536	0.0277092706815651\\
0.0132623574367439	0.0277211786601774\\
0.0132224119230675	0.0277413341315904\\
0.0131743170346945	0.0277760745183204\\
0.0131154312153793	0.0278387739725025\\
0.0130422721667354	0.0279620545232888\\
0.0129523819146622	0.0282476934940976\\
0.0128762310555204	0.0292243410158573\\
};
\addlegendentry{Fusion}

\end{axis}
\end{tikzpicture}%
    \vspace{-1.cm}
  \centerline{(b) Fusion} \medskip
\end{minipage}
\vspace{-0.4cm}
\caption{
Comparison of the tradeoff in terms of the square root of the \ac{crb} between \ac{bp} and \ac{ms} across various schemes: (1) Holistic perspective; (2) Closer examination of the fusion scenario.}
\label{tradeoff}
% \vspace{-0.2cm}
\end{figure}

\subsubsection{Tradeoff between \ac{bp} and \ac{ms}}
In Fig. \ref{tradeoff-diff-scheme}, we examine the performance tradeoff between \ac{bp} and \ac{ms}, measured by the square root of the \ac{crb}, across different target numbers and under various schemes. Target numbers range from one to three, as at least one target is required for feasible \ac{bp} in the considered setting \cite{henk2019twc}. All curves reveal the fundamental tradeoff between \ac{bp} and \ac{ms}. For digital schemes, we compare results across three paradigms: WCRB, WBF, and WCM. Among these, the WCRB approach demonstrates the most favorable bistatic-monostatic performance tradeoff, outperforming the weighted-sum mismatch approaches (WBF and WCM), as the latter exhibit weak Pareto frontier. Notably, within both Digital-\ac{fdb} and Digital-\ac{cpa} approaches, schemes based on WCM consistently show a superior bistatic-monostatic performance tradeoff compared to those based on WBF, with results nearly reaching the weak Pareto frontier\cite{ehrgott2005multicriteria}. Additionally, in the Digital-\ac{fdb} scheme, the performance gap between WBF and WCM becomes more pronounced as the target number increases. This observation suggests that approximating the covariance matrix directly preserves the desired spatial characteristics of the transmitted signal more effectively than beamformer approximation, as the \ac{fim} elements are determined directly by the covariance matrix. This insight supports the motivation for adopting the WCM paradigm in developing analog beamforming schemes.

It is also worth noting an interesting trend: as the number of targets increases, \ac{bp} \ac{crb} decreases, while \ac{ms} \ac{crb} increases, causing the tradeoff curves to shift upward and to the left with increasing $K$. This can be attributed to each resolvable target enhancing the position-related information of the \ac{ue}\cite{henk2019twc}, thus improving \ac{bp} performance. In contrast, a higher target number imposes a heavier load on \ac{ms} due to the limited, fixed spatial \acp{dof} that must be allocated across multiple targets, leading to reduced overall \ac{ms} performance.

In Fig. \ref{tradeoff}, we compare the performance tradeoff between \ac{bp} and \ac{ms} across various schemes for the case when $K=3$. Additionally, the Fusion scheme is presented to illustrate the advantages of information exchange between the \ac{bs} and \ac{ue}, enabling improved joint \ac{bp} and \ac{ms} performance. Fig. \ref{tradeoff}(a) offers a comprehensive comparison across all approaches, where we observe a noticeable performance reduction from the proposed digital approaches to their analog counterparts. This reduction is attributed to the limited design \acp{dof} in analog approaches, constrained by unit-modulus requirements necessary to maintain analog characteristics. Furthermore, within both digital and analog categories, the \ac{fdb} schemes significantly outperform \ac{cpa} in the bistatic-monostatic performance tradeoff, due to their higher optimization \ac{dof}. The Fusion scheme surpasses all other approaches by fully leveraging shared tasks between \ac{bp} and \ac{ms}, highlighting the considerable benefits of this mutualistic mechanism for enhancing joint \ac{bp} and \ac{ms} performance. Finally, we note that the Fusion scheme minimizes the tradeoff between \ac{bp} and \ac{ms} to such an extent that Fig. \ref{tradeoff}(a) displays it as nearly a single point. However, as shown in Fig. \ref{tradeoff}(b), even when minimized, a fundamental tradeoff between \ac{bp} and \ac{ms} persists, which becomes apparent upon closer inspection.

\section{Conclusion}
In this study, we investigated the joint optimization of \ac{bp} and \ac{ms} within a \ac{mimo} \ac{ofdm} framework, proposing innovative beamforming strategies that enable flexible tradeoff between \ac{bp} and \ac{ms} performance. We derived \acp{crb} for both \ac{bp} and \ac{ms}, targeting two primary objectives: allowing user equipment to estimate its position while accounting for clock bias and orientation mismatches, and enabling the \ac{bs} to localize passive targets. This led to a multi-objective optimization problem for beamforming. We analyzed digital schemes using weighted-sum \ac{crb} and mismatch minimization approaches, evaluating their performance under \ac{fdb} and \ac{cpa}. To enhance hardware efficiency, we developed analog \ac{fdb} and \ac{cpa} strategies based on covariance matrix mismatch minimization. Our numerical results demonstrate the effectiveness of the proposed designs, highlighting the advantages of minimizing covariance matrix mismatch and underscoring the benefits of information fusion between \ac{bp} and \ac{ms} for practical system implementation. 
There are several interesting avenues for future research, including: 1) considering the uncertainty in \ac{ue}/target positions, 2) examining different visibility conditions between \ac{bp} and \ac{ms} in terms of targets, and 3) analyzing targets in the near field of \ac{bs}/\ac{ue}.

\appendices

\section{Fundamentals of the \ac{sqp} Framework}\label{sqp-fundamental}
According to \cite{wright2006opt}, the \ac{sqp} algorithm is an iterative approach for addressing nonlinear optimization problems, formulated as
\begin{subequations}\label{problem-describe}  
\begin{align}  
\mathop {\min }\limits_{\boldsymbol{x}} \;\;\;  
& f\left(\boldsymbol{x}\right) \label{problem-describe-obj}\\  
{\rm{s.t.}}\;\;\;  
& g_i\left(\boldsymbol{x}\right) = 0, \quad i = 1, \ldots, m, \\
& h_j\left(\boldsymbol{x}\right) = 0, \quad j = 1, \ldots, p, 
\end{align}  
\end{subequations}  
where $f(\boldsymbol{x})$ represents the objective function, while $g_i(\boldsymbol{x})$ and $h_j(\boldsymbol{x})$ denote the equality and inequality constraints, respectively.

The primary components integral to understanding the \ac{sqp} workflow are as follows.
\begin{enumerate}
\item \emph{Lagrangian Function:} Defined as $\mathcal{L}(\boldsymbol{x}, \boldsymbol{\lambda}, \boldsymbol{\mu}) = f(\boldsymbol{x}) + \sum_{i=1}^{m} \lambda_i g_i(\boldsymbol{x}) + \sum_{j=1}^{p} \mu_j h_j(\boldsymbol{x}),$ where $\boldsymbol{\lambda} = [\lambda_1,\ldots,\lambda_m]^{\mathsf{T}}$ and $\boldsymbol{\mu} = [\mu_1,\ldots,\mu_p]^{\mathsf{T}}$ are Lagrange multipliers for equality and inequality constraints.
\item \emph{Quadratic Subproblem:} At the $k$-th iteration, \ac{sqp} solves a quadratic approximation of the Lagrangian around $\boldsymbol{x}_k$ under linearized constraints, expressed as
\begin{subequations}\label{sqp-subproblem}  
\begin{align}  
\mathop {\min }\limits_{\boldsymbol{p}} \;\;\;  
& \nabla f\left(\boldsymbol{x}_k\right)^{\mathsf{T}} \boldsymbol{p} + \frac{1}{2} \boldsymbol{p}^{\mathsf{T}} \nabla^2 \mathcal{L}\left(\boldsymbol{x}_k, \boldsymbol{\lambda}_k, \boldsymbol{\mu}_k\right) \boldsymbol{p} \label{sqp-subproblem-obj}\\  
{\rm{s.t.}}\;\;\;  
& \nabla g_i\left(\boldsymbol{x}_k\right)^{\mathsf{T}} \boldsymbol{p} + g_i\left(\boldsymbol{x}_k\right) = 0, \quad \forall i, \\
& \nabla h_j\left(\boldsymbol{x}_k\right)^{\mathsf{T}} \boldsymbol{p} + h_j\left(\boldsymbol{x}_k\right) \leq 0,  \quad \forall j,
\end{align}  
\end{subequations}  
where $\nabla f(\boldsymbol{x}_k)$ and $\nabla^2 \mathcal{L}(\boldsymbol{x}_k, \boldsymbol{\lambda}_k, \boldsymbol{\mu}_k)$ are the gradient vector and Hessian matrix of $f(\boldsymbol{x})$ evaluated at $\boldsymbol{x} = \boldsymbol{x}_0$.
\item \emph{Merit Function:} To ensure simultaneous improvement in the objective and constraints, a merit function is defined as $\phi(x, \lambda) = f(x) + \alpha ( \sum_{i=1}^{m} |g_i(x)| + \sum_{j=1}^{p} \max(0, h_j(x)) )$ with $\alpha$ as a penalty parameter.
\end{enumerate}

The \ac{sqp} algorithm follows these steps.
\begin{enumerate}
\item \emph{Initialization:} Choose an initial guess $\boldsymbol{x}_0$ and set initial multipliers $\boldsymbol{\lambda}_0$ and $\boldsymbol{\mu}_0$.
\item \emph{Quadratic Subproblem:} At the $k$-th iteration, solve \eqref{sqp-subproblem} to determine the search direction $\boldsymbol{p}_k$.
\item \emph{Line Search or Trust Region:} Select a step size $\overline{\beta}_k$ that ensures a sufficient decrease in the merit function.
\item \emph{Update:} Set $\boldsymbol{x}_{k+1} = \boldsymbol{x}_k + \overline{\beta}_k \boldsymbol{p}_k$ and update $\boldsymbol{\lambda}_k$ and $\boldsymbol{\mu}_k$.
\item \emph{Check:} Repeat until the Karush–Kuhn–Tucker conditions are approximately met.
\end{enumerate}

\bibliographystyle{IEEEtran}
\bibliography{IEEEabrv,mybib}

\end{document}